\documentclass[12pt]{iopart}

\usepackage{cite}
\usepackage{graphicx}
\graphicspath{ {./figures/} }
\usepackage{epsfig}
\usepackage{epstopdf}
\usepackage{textcomp}
\usepackage{multirow}
\usepackage{setspace}
\usepackage{fullpage}
\usepackage{float}
\usepackage{caption}
\usepackage{subcaption}
\usepackage{iopams}  
\usepackage{siunitx}
\begin{document}

\title{On the primary spacing and microsegregation of cellular dendrites in laser deposited Ni-Nb alloys}

\author{Supriyo Ghosh$^1$\footnote{Corresponding author. Email address: supriyo.ghosh@nist.gov. }, Li Ma$^1$ $^2$, Nana Ofori-Opoku$^1$ $^3$ and Jonathan~E.~Guyer$^1$ }

\address{$^1$ Materials Science and Engineering Division, National Institute of Standards and Technology, Gaithersburg, MD 20899, USA}
\address{$^2$ Theiss Research, La Jolla, CA 92037, USA}
\address{$^3$ Center for Hierarchical Materials Design, Northwestern University, Evanston, IL 60208, USA}
\vspace{10pt}

\begin{abstract}
In this study, an alloy phase-field model is used to simulate solidification microstructures at different locations within a solidified molten pool. The temperature gradient $G$ and the solidification velocity $V$ are obtained from a macroscopic heat transfer finite element simulation and provided as input to the phase-field model. The effects of laser beam speed and the location within the melt pool on the primary arm spacing and on the extent of Nb partitioning at the cell tips are investigated. Simulated steady-state primary spacings are compared with power law and geometrical models. Cell tip compositions are compared to a dendrite growth model. The extent of non-equilibrium interface partitioning of the phase-field model is investigated. Although the phase-field model has an anti-trapping solute flux term meant to maintain local interface equilibrium, we have found that during simulations it was insufficient at maintaining equilibrium. This is due to the fact that the additive manufacturing solidification conditions fall well outside the allowed limits of this flux term.

\end{abstract}
\section{Introduction}
Demand for improved strength and resistance to creep at elevated temperatures makes Ni alloys suitable for aerospace and other industrial applications~\cite{Reed2008}. Recently, the laser powder bed fusion (L-PBF) additive manufacturing process has been introduced in which near-net metallic objects are produced from powder, in a layer by layer fashion with repeated solidification and solid-state phase transformations, in a shorter manufacturing time and with almost no finishing~\cite{Murr2012}. Solidification in this process controls the size and shape of the grains, the growth morphology, the extent of microsegregation, and ultimately the properties of the product. Therefore, understanding of the melt pool solidification behavior is essential.

As the laser rasters across the powder bed, local regions begin to melt resulting in a molten pool of certain dimensions (width and depth). The molten pool then undergoes a solidification process driven by the resultant complex thermal history due to repeated passes of the laser. Solidification begins along the back side of the moving melt pool boundary at a location defined by the liquidus temperature of the given alloy. The microstructures will vary, controlled by the local temperature gradient $G$, and solidification rate $V$ and in turn the cooling rate ($\dot{T}$ or $GV$). A low $V$ and high $G$ exist at the bottom of the melt pool whereas a higher $V$ and lower $G$ exist close to the melt pool surface. Moreover, in the solidification environment generated by the laser melting process, $V$ is found to be far more important than $G$~\cite{Kurz1994}. With this motivation, two primary components of cellular/dendritic growth are illustrated: primary arm spacing and microsegregation.

For additive manufacturing, materials strength is related to the cell/dendrite spacing and microsegregation. There have been several studies performing experiments~\cite{Formenti2005,Murr2012,Harrison2015,gaumann2001,Whitesell2000, Amato2012} and simulations~\cite{Lee2014,Liang2016,nie2014,Wang2003,Lee2010,Trevor2017} regarding microstructural features in additive Ni-Nb alloys. However, the primary arm spacing and the microsegregation across the cells/dendrites have not received much attention. In addition, the solidification conditions, $G$, $V$ and $\dot{T}$, considered in prior work are relatively small-valued, whereas the present work treats larger values of the above solidification conditions to be applicable to additive manufacturing. Therefore, the characteristics of the parameter-microstructure map are expected to be different than those reported in the existing literature. Many theoretical approaches have been developed over the past decades on the basis of a predefined shape of an isolated cell/dendrite to estimate the primary arm spacing $\lambda_1$ with the solidification parameters~\cite{Hunt1979book,kurzbook}: $\lambda_1 = A G^mV^n$, where $A$ is an alloy constant and $m$ and $n$ are the model-dependent exponents. However, in additive manufacturing applications, these approximations are less valid due to the complex nature of the diffusion fields around the cells/dendrites at high solidification rates. Despite these limitations, the present work is compared with these theoretical estimates for reference.

During solidification, solute gets partitioned between solid and liquid phases to reach the equilibrium compositions corresponding to the phase diagram. Interestingly, while at low velocities the above partitioning is in equilibrium, at intermediate to high velocities solute redistribution is not complete and thus solute compositions across interface do not follow the phase diagram~\cite{Boettinger1999,kurz1986,amberg2008}. Microstructures presented in this work are for intermediate to rapid cooling rates with the solidification velocities more than 1 cm s$^{-1}$~\cite{Kurz1994}. Therefore, to reflect this in the processing-microstructure map, solidification parameters are modified from equilibrium to velocity-dependent values~\cite{Kurz1994,Liang2016}. Microstructure evolution with non-equilibrium interface partitioning has received little attention, but it is important to consider for the laser powder deposition process~\cite{Kurz1994,Liang2016}.

Due to the high temperature and small volume of the molten pool, \textit{in situ} measurements of the solidification conditions are difficult. Numerical simulations of the laser deposition process is a viable alternative to obtain local solidification conditions. For this purpose, a 3-D heat transfer Finite Element Analysis (FEA) code is used to obtain melt pool shapes as well as temperatures. For the microstructure simulations, a phase-field method is employed. This method has become a popular choice over the past few decades to study solidification problems\cite{chen2002, boettinger2002,steinbach2009, moelans2008, supriyo2014, Ghosh2015}. In this method, a scalar-valued order parameter field $\phi$ is introduced to distinguish the constituent phases in a microstructure: if liquid is defined by $\phi =$ -1 and solid by $\phi = $1, then the (diffuse) interface is described by $-1<\phi<1$, and the position of the interface can be taken as the $\phi = 0$ contour. Therefore, explicit tracking of the interface is no longer needed, and we can simulate complex liquid-solid interfaces in an efficient way. Despite the advantages of the phase-field method, it still requires considerable computation time.

The objective of this study is twofold. First, to perform FEA simulations to determine actual solidification conditions in the melt pool and, second, to perform phase-field simulations using the above conditions to calculate the sizes and the concentrations of the dendrite cells for Ni-Nb, a binary approximation of a Ni-based superalloy. In section~\ref{sec_methods}, we describe the FEA and phase-field simulation methods, parameters, and procedures. In section~\ref{sec_results}, FEA results are analyzed to obtain the local solidification conditions and used for phase-field solidification microstructure predictions. In the phase-field results, primary arm spacing, concentration profile around the tip, and the extent of microsegregation are determined. Finally, a summary is outlined along with a possible outlook in section~\ref{sec_summary}. 

\section{Simulation Methods}\label{sec_methods}
A macroscopic heat transfer code is used to simulate melt pool thermal history profiles. Subsequently, a phase-field method is used to simulate solidification microstructures under the obtained thermal profiles.

\subsection{Macroscopic heat transfer model}
Using the commercial FEA code ABAQUS~\cite{Abaqus},~\footnote{Certain commercial equipment, instruments, or materials are identified in this paper in order to specify the experimental procedure adequately. Such identification is not intended to imply recommendation or endorsement by NIST, nor is it intended to imply that the materials or equipment identified are necessarily the best available for the purpose.} a non-linear, transient, thermal model is designed and implemented to obtain the global temperature history generated during laser irradiation of one layer of powder deposited on a solid substrate. Inconel 718 alloy (IN718) is used for the powder as well as the substrate properties in the simulation. A single-track laser scan across a layer of loose metal powder with thickness \SI{36}{\micro\metre} is modeled. To reduce computational time, the elements that interact with the laser beam are finely meshed within the diameter of the laser beam, and a coarse mesh is used for the surrounding loose powder and the substrate (refer to figure~\ref{fig_3d_mp}). The temperature distribution $T$ throughout the substrate and the powder, as a function of time $t$, is determined by solving a transport equation for thermal energy as follows~\cite{Bird1960}:
\begin{equation}\label{eq_conduction}
\nabla \cdot (\kappa\nabla T) + Q = \frac{\partial(\rho c_p T)}{\partial t},
\end{equation}
where $\kappa$ is the thermal conductivity, $Q$ is the heat source (heat of melting/solidification), $c_p$ is the specific heat capacity and $\rho$ is the density. The initial condition assumes a uniform temperature of 352 K in the powder and the substrate at time $t = 0$. For the boundary conditions, heat input from the laser $q_s$, heat loss via external convection and radiation are considered at the top surface via 
\begin{equation}\label{eq_radiation_convection}
\kappa (-\nabla T \cdot \hat{N}) = q_s + h(T-T_a) + \epsilon_R \sigma_R(T^{4}-T_{a}^{4}).
\end{equation}
Here $T_a$ is the ambient temperature, $\hat{N}$ the vector normal to the surface, $h$ the convective heat transfer coefficient, $\epsilon_R$ the thermal radiation coefficient and $\sigma_R$ the Stefan-Boltzmann constant.

The transient thermal profile is computed by ABAQUS based on the material properties input and the boundary and initial conditions applied. In this case, heat input on the top surface is simulated using a surface heat source moving along the $x$ direction obeying a Gaussian distribution~\cite{Roberts2009},
\begin{equation}\label{eq_gauss}
q_s = \frac{2 \eta P}{\pi r_{b}^2}\exp\left(-\frac{2r^2}{r_b^2}\right),
\end{equation}  
where $P$ is the total laser power, $\eta$ is the power absorption coefficient, $r_b$ is the laser beam radius and $r$ is the radial distance to the beam centerline. The values of these parameters are listed in table~\ref{table_param_laser}. Other types of heat source models have also been used to describe the laser melting process, such as moving point, line or plane heat sources proposed by Rosenthal~\cite{Rosenthal1941, Rosenthal1946}, or a double ellipsoidal volumetric heat source proposed by Goldak \emph{et al.}~\cite{Goldak1984}. It should be noted that those heat source equations can also be solved analytically using the same laser processing parameters to predict the temperature distribution in a molten pool~\cite{Rosenthal1946, Goldak1984, Eagar1983, Nguyen1999}. Moreover, since the shape of the simulated melt pool is not as complex as in keyhole mode melting and the layer thickness of powders is very small (\SI{36}{\micro\metre}), a Gaussian distribution can be used to describe the melt pool shape approximately, as suggested in~\cite{Guo2000,Suresh2016}.

\begin{table}[h]
\caption{Laser processing parameters used in the FEA simulations.}\label{table_param_laser}
\centering
\begin{tabular}{ll}
\hline
Laser Power, $P$ &  195 W	\\
Laser scan speed, $V$ &	0.8 m s$^{-1}$ \\
Laser beam radius, $r_b$ &	\SI{50}{\micro\metre}	\\
Absorption coefficient, $\eta$ &	0.5			\\
\hline
\end{tabular}
\end{table}

The overall powder quality depends on the packing density and particle characteristics, such as size distributions, size ratios and surface morphologies, which in turn determine the morphology and properties of the final additively manufactured parts~\cite{Santomaso2003}. The initial powder packing density, given by the ratio of local density of the powder to the density of the bulk solid, is estimated as $50$\% for the present simulations. Powder density linearly increases from powder to bulk values as the $T$ rises from the solidus temperature $T_s$ to the liquidus temperature $T_l$ given by the IN718 phase diagram predicted by CALPHAD-based thermodynamic calculations~\cite{TCNI,Andersson2002}. Above $T_l$, the initial powder-state elements are irreversibly changed to bulk-state elements. Thermal conductivity $\kappa$ of the powder bed depends not only $T$, but also on the packing fraction, particle size distribution, particle morphology, and thermal conductivity of the bulk material and surrounding gas~\cite{Childs2005, Tanaka2012}. As a first approximation, the present work treats $\kappa$ as a function of $T$ only. All the above changes are performed by the ABAQUS user subroutine. For details, please refer to~\cite{Brandon2015,LiMa2015}.

The CALPHAD approach can be used to accurately predict thermodynamic properties, particularly for additive manufacturing applications, by studying the solidification behavior of multicomponent alloys. The $T$-dependent bulk material density, latent heat and specific heat are calculated under local equilibrium conditions by feeding the nominal IN718 powder composition as an input to the built-in module of the Thermo-Calc software~\cite{Andersson2002} which uses the TCNI version 8 thermodynamic database~\cite{TCNI}. This database is generally considered accurate for Ni-based superalloys research. A recent review on coupling of CALPHAD approach to 3-D FEA can be found in~\cite{Smith2016_1}.

L-PBF is a complex process where a wide range of transient non-equilibrium physical phenomena take place within the molten pool. Some of these factors are viscous forces, buoyancy forces, melt convection, Marangoni convection, evaporation cooling and recoil pressure~\cite{Khairallah2016,King2015}. The present model ignores the above physics for simplicity. We do not consider the formation of topographic depressions in the melt pool as in keyhole mode melting, which appears due to the spatter ejection during deep penetration of the incident very high power laser beam~\cite{King2014}. We ignore any formation of oxide layers in the powder particles. For a system with low to moderate power density of the laser beam and very small thickness of powder layers, the underlying assumptions may be appropriate, as suggested in~\cite{Guo2000,Suresh2016}. However, for accurate modeling of the melt pool, the above multiphysics need to be included, although it may be computationally expensive.

\subsection{Mesoscopic phase-field model}
For the phase-field simulations, we use a quantitative alloy phase-field model presented by Echeberria \emph{et al.} \cite{Echebarria2004}. This model is formulated in the thin-interface limit to remove the interface thickness dependencies resulting in a faithful description of the (non-conserved) phase-field $\phi$ and the (conserved) composition field $c$ during solidification of a dilute binary alloy. An anti-trapping current was introduced~\cite{Karma2001, Echebarria2004} in this model to avoid spurious solute-trapping effects arising from thick diffuse interfaces at low solidification velocities. However, as shown below, the model does not prevent solute-trapping at high velocities. The effects of convection are not considered in this model and thus the solute is transported in the liquid by diffusion only. Referring to the fields $\phi$ and $c$, the equations of the model in 2-D are given by

\begin{eqnarray}\label{eq_phi}
\fl \tau_0 a(\theta)^2\frac{\partial \phi}{\partial t} = W_{0}^{2} \nabla \cdot \left[{a(\theta)}^2 \nabla\phi\right] - \frac{\partial}{\partial x} \left[ a(\theta) a^{'}(\theta) \frac{\partial \phi}{\partial y} \right] + \frac{\partial}{\partial y} \left[ a(\theta) a^{'}(\theta) \frac{\partial \phi}{\partial x} \right] \nonumber \\
+\phi -\phi^3 - \frac{\lambda}{1-k_e} (1-\phi^2)^2 \left[e^u -1 + \frac{T- T_0}{m_l c_0/k_e}\right],
\end{eqnarray}

and

\begin{equation}\label{eq_c}
\fl \frac{\partial c}{\partial t} = \nabla \cdot \left( \left[\frac{1-\phi}{2} {D_l} + \frac{1+\phi}{2} {D_s}\right] \lbrace 1+k_e-(1-k_e)\phi\rbrace \frac{c_0}{k_e} \nabla e^u + \frac{W_0}{2\sqrt{2}} (1-k_e) \frac{c_0}{k_e} e^u \frac{\partial \phi}{\partial t} \hat{n}\right).
\end{equation}

Equation~\ref{eq_phi} couples bulk thermodynamics with interface effects. $a(\theta) = 1 + \epsilon \cos (4\theta$) represents the two-dimensional fourfold anisotropy at the solid-liquid interfaces with strength $\epsilon$, $\theta = \arctan\left(\partial_y\phi/\partial_x\phi\right)$ is the angle between the interface normal and the $x$ direction in the lab frame of reference. $a^{'}(\theta)$ denotes the derivative of $a(\theta)$ with respect to $\theta$. $c_0$, $m_l$ and $k_e$ are thermophysical properties of the material and, for a dilute alloy, they are nominal composition, liquidus slope and equilibrium partition coefficient, respectively. $k_e=c_s/c_l$, where $c_s$ and $c_l$ are the equilibrium compositions on the solid and liquid side of the interface, respectively. $u$ is a dimensionless chemical potential given by: $\ln \left(\frac{2ck_e/c_0}{1+k_e-(1-k_e)\phi}\right)$. A frozen-temperature approximation is applied in which an imposed temperature gradient $G$ is translated along the $y$ (growth) axis with a constant speed $V$ following $T = T_0 + G(y-Vt)$, where $T_0(y=0, t=0)$ is a reference temperature.

In equation~\ref{eq_c}, the first term inside parentheses represents a Fickian diffusion flux and the second term is the anti-trapping current. $D_s$ and $D_l$ are the diffusivity of solute in solid and liquid, respectively. $\hat{n} =\nabla\phi/|\nabla \phi|$ is the unit vector normal to the interface. 

There are three characteristic parameters in this model, $W_0$: the interface thickness, $\tau_0$: the phase-field relaxation time, and $\lambda$: a dimensionless coupling constant. These parameters are linked to the physical quantities by two relations: one via chemical capillary length $d_0 = a_1 W_{0}/\lambda$ and the other via setting the interface kinetics to zero to maintain the local equilibrium at the interface yielding $\tau_0 = a_2\lambda W_{0}^{2}/D_l$. The constants $a_1$ and $a_2$ are given by $a_1$ = 0.8839 and $a_2$ = 0.6267~\cite{Karma2001}. This way, $W_0$ becomes the only free parameter which is chosen depending on the scale of the simulated microstructures. Although we are assuming negligible interface kinetics, we note that in the intermediate to rapid solidification regime, we cannot expect such an assumption to be completely valid. At present, the authors are not aware of any atomistic level simulations that can offer realistic values of the interface kinetic coefficients for Ni-Nb or any Ni-based superalloys. As a first approximation to understand and examine the consequences of that assumption on the microstructure evolution in these alloys, we assume zero interface kinetics.

\subsection{Simulation procedures}
For the phase-field simulations, phase-field (equation~\ref{eq_phi}) and concentration (equation~\ref{eq_c}) equations of motion are solved on a uniform mesh, using a finite volume method and an explicit time marching scheme. Zero-flux boundary conditions are applied on both $\phi$ and $c$ fields in all directions. The size of the simulation box in the ($y$) growth direction is taken as \SI{40}{\micro\metre}, which is at least 150 times the diffusion length $D_l/V$, and varying domain sizes are used in the $x$-direction ranging from \SI{5}{\micro\metre} to \SI{10}{\micro\metre} depending on the fineness of the simulated cellular structures. For each simulation, a grid spacing of $\Delta x$ = $\Delta y$ = \SI{0.008}{\micro\metre} is used. A maximum time-step of $\Delta t$ = \SI{0.003}{\micro\second} is found to be numerically stable for the present calculations. The maximum phase-field interface thickness $W_0$ used is \SI{0.01}{\micro\metre}, yielding $\lambda$ = 1.377. Note that this value of $W_0$ is roughly 10 times smaller than the dendrite tip radius calculated by a sharp-interface model for the same parameters. 

Each simulation is initialized with a thin layer of solid of height \SI{0.05}{\micro\metre} from the bottom of the simulation box with an initial Nb composition of $k_ec_0$ in the solid and $c_0$ in the liquid. Random, small amplitude perturbations are applied at the initial solid-liquid interface. Stable perturbations have been found to grow with time and break into steady-state cellular structures. At this stage, cell tips grow at a constant temperature and at a constant velocity equal to the solidification velocity. It should be noted that we have approximated the alloy IN718 to be a binary Ni-\SI{5}{\%} Nb~\footnote{Concentration is represented in mass fraction in the present paper.} in this study and the corresponding quasi-binary phase diagram has nearly constant liquidus slope $m_l$ as well as partition coefficient $k_e$~\cite{knorovsky1989}. The possible formation of Ni$_3$Nb in the intercellular regions is not treated in this paper. The thermophysical parameters of the dilute Ni-Nb alloy are taken from Nie \emph{et al.}~\cite{nie2014} and listed in table~\ref{table_param_pf}. The processing parameters, $G$ and $V$ in equation~\ref{eq_phi}, are extracted from the melt pool solid-liquid boundary given by the FEA simulations. This is further detailed in the following section. 

\begin{table}[h]
\caption{Material properties used in the simulations, after~\cite{nie2014}.}\label{table_param_pf}
\centering
\begin{tabular}{ll}
\hline
Initial alloy mass fraction, $c_0$ 		&	\SI{5}{\%} \\
Equilibrium Partition Coefficient, $k_e$	&	0.48	\\
Liquidus Slope, $m_l$	&	-10.5 K \%$^{-1}$	\\
Equilibrium Freezing Range, $\Delta T_0$ = $T_l - T_s$ 		&	57 K	\\
Liquid Diffusion Coefficient, $D_l$	& $3 \times 10^{-9}$ m$^2$ s$^{-1}$ \\
Solid Diffusion Coefficient, $D_s$	& $10^{-12}$ m$^2$ s$^{-1}$	\\
Anisotropy Strength, $\epsilon$	& \SI{3}{\%}	\\
Capillary Length, $d_0$ & $8.0 \times 10^{-9}$ m \\
Gibbs-Thomson coefficient, $\Gamma$ &  $3.65 \times 10^{-7}$ K m \\
\hline
\end{tabular}
\end{table}%

\section{Results and Discussion}\label{sec_results}
First, we present the macroscopic heat transfer simulation results to illustrate the temperature distributions and to estimate the local solidification conditions ($G$ and $V$) at different positions along the melt pool boundary. Following this, the local cellular patterns at these locations are presented using phase-field simulations.
\subsection{FEA simulations: Estimation of $G$ and $V$}\label{sec_results_fea}
\begin{figure}[h]
\begin{center}
\includegraphics[scale=0.15]{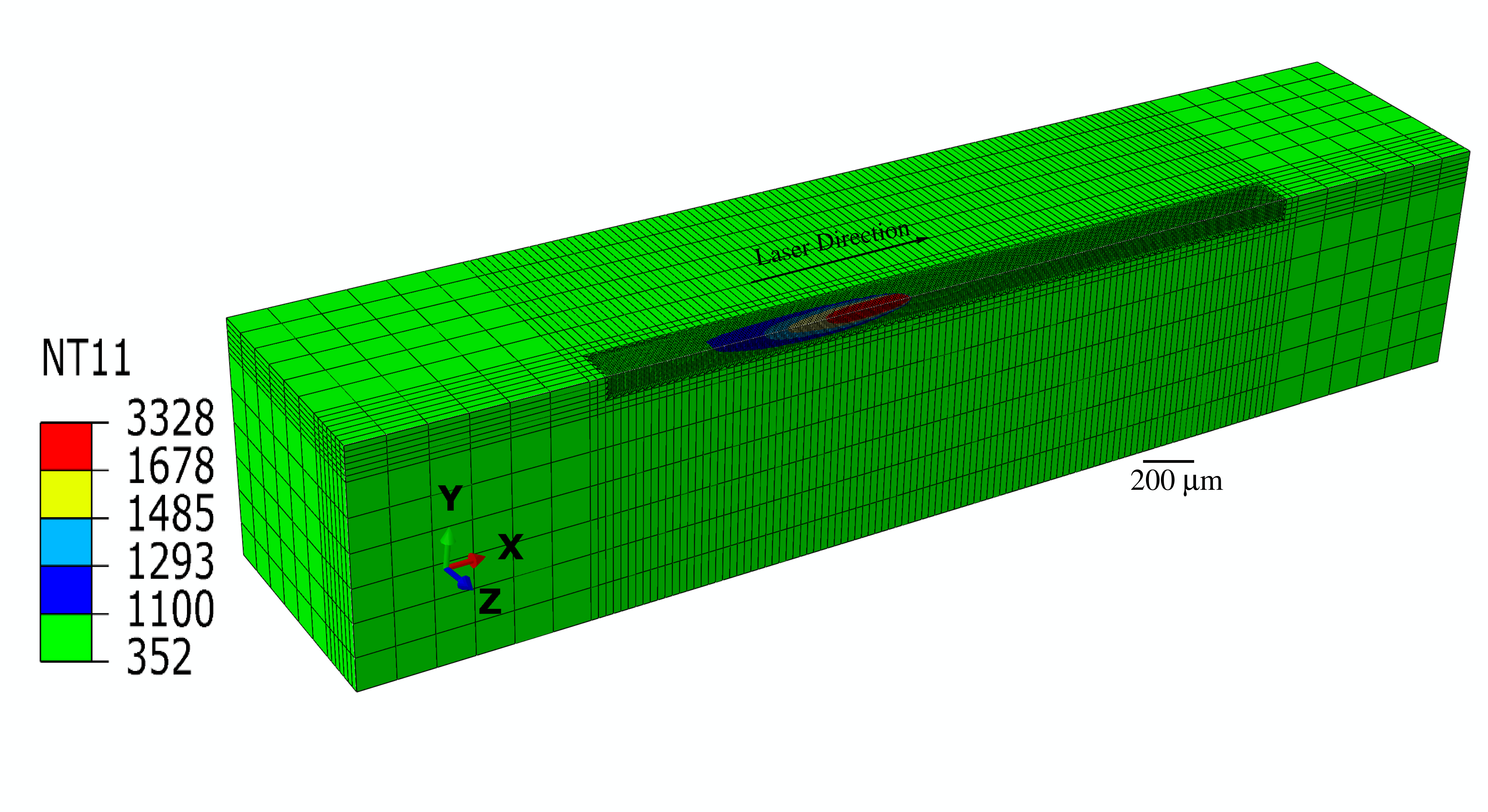}
\caption{FEA simulated 3-D melt pool shape for the parameters in table~\ref{table_param_laser}. Here, different colors represent different temperature zones, which are measured in Kelvin. Red represents the liquid melt pool. In the FEA simulations, a specimen geometry of $\SI{5}{mm}\times \SI{2}{mm} \times \SI{2}{mm}$ is used. A mesh of $\SI{10}{\micro\metre} \times\SI{10}{\micro\metre} \times \SI{6}{\micro\metre}$ is used to discretize the elements which interact with the laser, with coarser discretization in the far field.}\label{fig_3d_mp}
\end{center}
\end{figure}

\begin{figure}[h]
\begin{center}
\includegraphics[scale=0.25]{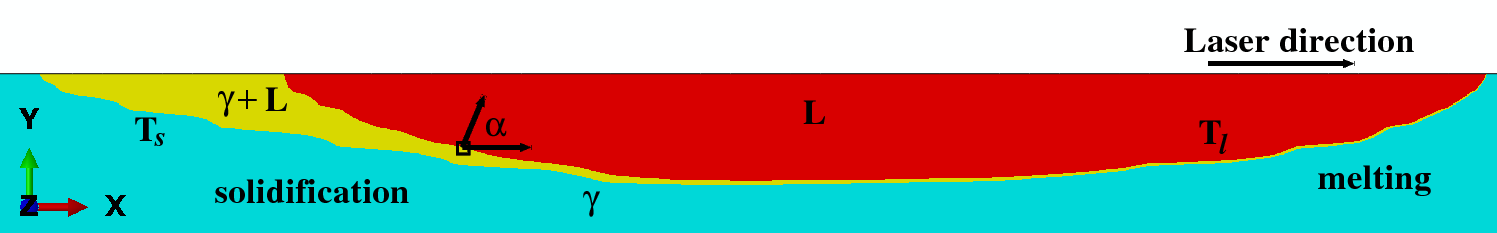}
\caption{A 2-D section cut along the centerline of the 3-D melt pool from figure~\ref{fig_3d_mp}. To correspond with the Ni-\SI{5}{\%} Nb phase diagram~\cite{knorovsky1989}, red represents the liquid phase (L), yellow represents solid and liquid coexistence (mushy zone) and cyan represents the solid (fcc) $\gamma$ phase. The solidification temperature isotherm is given by the boundary between red and yellow. The local solidification conditions are estimated along this isotherm. It should be noted that color scales are different in figure~\ref{fig_3d_mp} and figure~\ref{fig_meltpool}. While $T_l = 1678$ K exactly corresponds to the red-yellow boundary of figure~\ref{fig_3d_mp}, $T_s = 1621$ K is interpolated from the yellow band in figure~\ref{fig_3d_mp}.} \label{fig_meltpool}
\end{center}
\end{figure}

In FEA simulations, the laser surface treatment processing parameters (refer to table~\ref{table_param_laser}) determine the solidification processing parameters $G$ and $V$ for a given alloy. While $G$ depends on the temperature profiles generated by the laser beam, $V$ depends on the melt pool geometry and the laser beam speed. The trailing edge of the melt pool is the solidification interface. Its location is approximated by the location of the liquidus isotherm (refer to figures~\ref{fig_3d_mp} and \ref{fig_meltpool}). In experimental microstructures, these isotherms are represented by the cell/dendrite tips, which grow perpendicular to the solid-liquid interface at a rate $V$. This provides the microscopic link between the laser beam speed $V_b$ with the local solidification speed $V$. Moreover, it is evident in figure~\ref{fig_meltpool} that the shape of the melt pool is rarely symmetric but tear-shaped. Due to the above geometrical requirements, $V$ does not equal $V_b$ and varies along the solidification boundary. $V$ increases rapidly from zero at the bottom of the melt pool to a larger value at the rear of the melt pool interface following the equation:  $V = V_b \cos \alpha$, where $\alpha$ is the solidification angle measured between the normal of the solid-liquid interface and the direction of laser travel. 

$G$ also varies along the solidification boundary from the bottom to the top of the pool. For a particular position along this boundary, $G$ is estimated by the magnitude of the gradient along the Cartesian directions, $G$ = $|\nabla T|$ = $\sqrt{(\partial_{x} T)^{2}+(\partial_{y} T)^{2}+(\partial_{z} T)^{2}}$, using the available temperature values from the neighboring elements. Therefore, the solidification boundary represents different $G$ and $V$. Thus the microstructure within the solidified puddle also varies with depth. We choose multiple microscopic volume elements along this boundary with estimated $G$ and $V$ values, to be used for the following phase-field simulations. We note that $V$ varies between 0.01 m s$^{-1}$ ($\alpha$ = 89$^\circ$) to 0.3 m s$^{-1}$ ($\alpha$ = 68$^\circ$) along the melt pool boundary. Similarly, $G$ varies between $ 2.4 \times 10^7$ K m$^{-1}$ at the bottom to $0.14 \times 10^7$ K m$^{-1}$ at the top of the melt pool boundary. It should be noted that the magnitudes of $V$ used in this work are in the range of rapid solidification processing and, in this parameter space, $G$ plays a minor role in the microstructure selection process~\cite{Kurz1994, Harrison2015}. Therefore, we employ a fixed value of $G = 10^7$ K m$^{-1}$ for the phase-field simulations.

\subsection{Phase-field simulations}\label{sec_results_pf}
\subsubsection{General features}\label{sec_general_features}
It is well documented that for a given alloy composition or $c_0$, $G$ and $V$ control the interface morphology as well as the scale of the solidification microstructures. There exist several criteria to examine if the evolving solidification interface will be planar or dendritic. For a given alloy composition, the lower limit of this transition can roughly be estimated by satisfying the constitutional supercooling criterion: $V_{cs} = \Delta T_0 / (G D_l)$ . The upper limit is given by the absolute stability criterion~\cite{Mullins1964}: $V_{ab} = \Delta T_0 D_l / (k_e \Gamma$). The physical meaning is that as long as $V$ is below $V_{cs}$, the interface will grow as a planar front which further breaks into cells/dendrites with increasing $V$, and for $V$ $>$ $V_{ab}$, cells/dendrites will be lost with re-establishment of the planar front. Thus, the essential interfacial features in the microstructures are governed by the solidification velocities. Note that the solidification velocities estimated from the FEA simulations are between $V_{cs}$ = 0.0005 m s$^{-1}$ $< V < V_{ab}$ = 1 m s$^{-1}$.  

\begin{figure}[h]
\centering
\begin{subfigure}[t]{0.55\textwidth}
\centering
\includegraphics[scale=0.2]{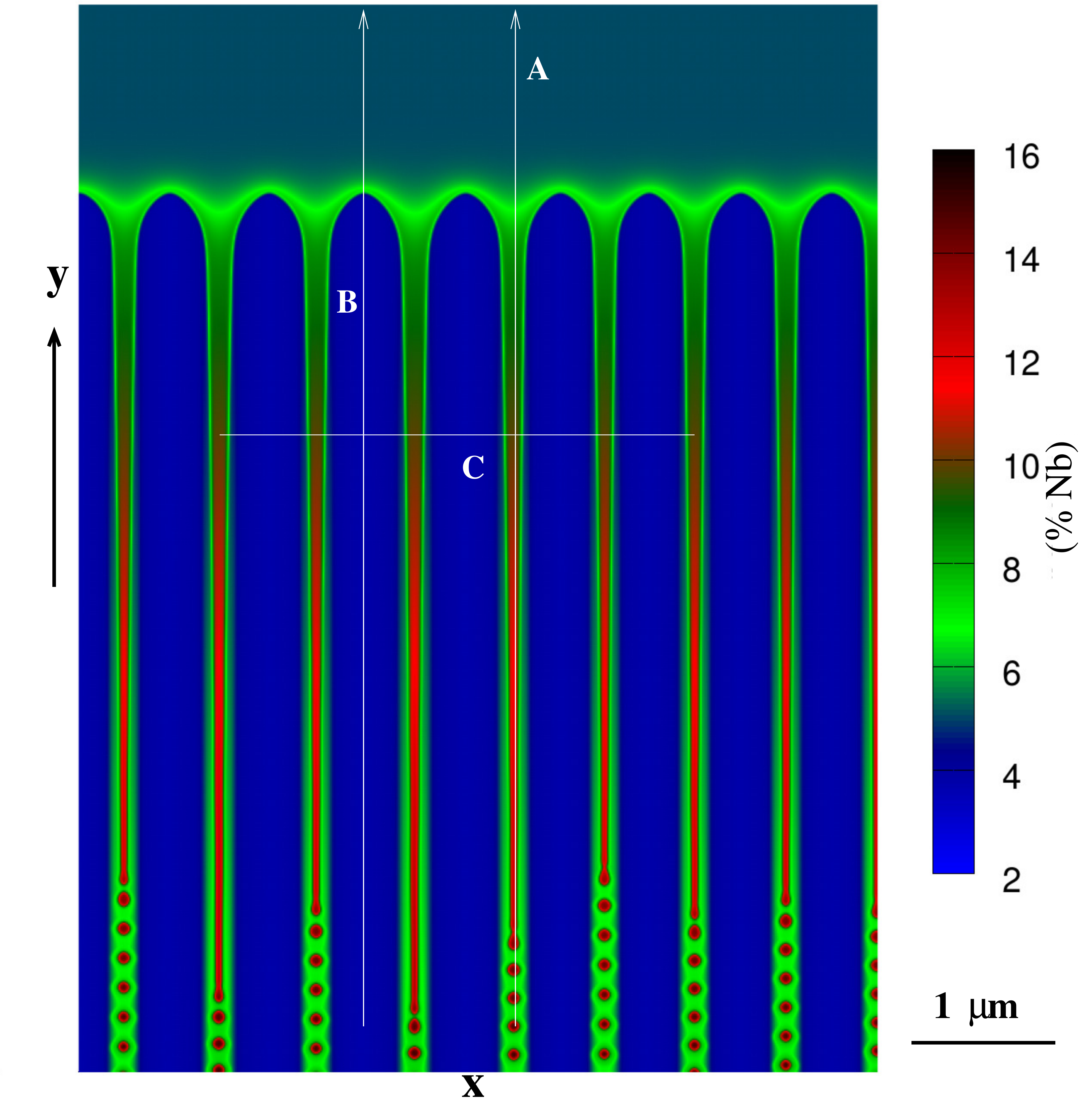}
\caption{}\label{fig_figure1}
\end{subfigure}%
\begin{subfigure}[t]{0.55\textwidth}
\centering
\includegraphics[trim={4cm 0 0 0},clip,scale=0.2]{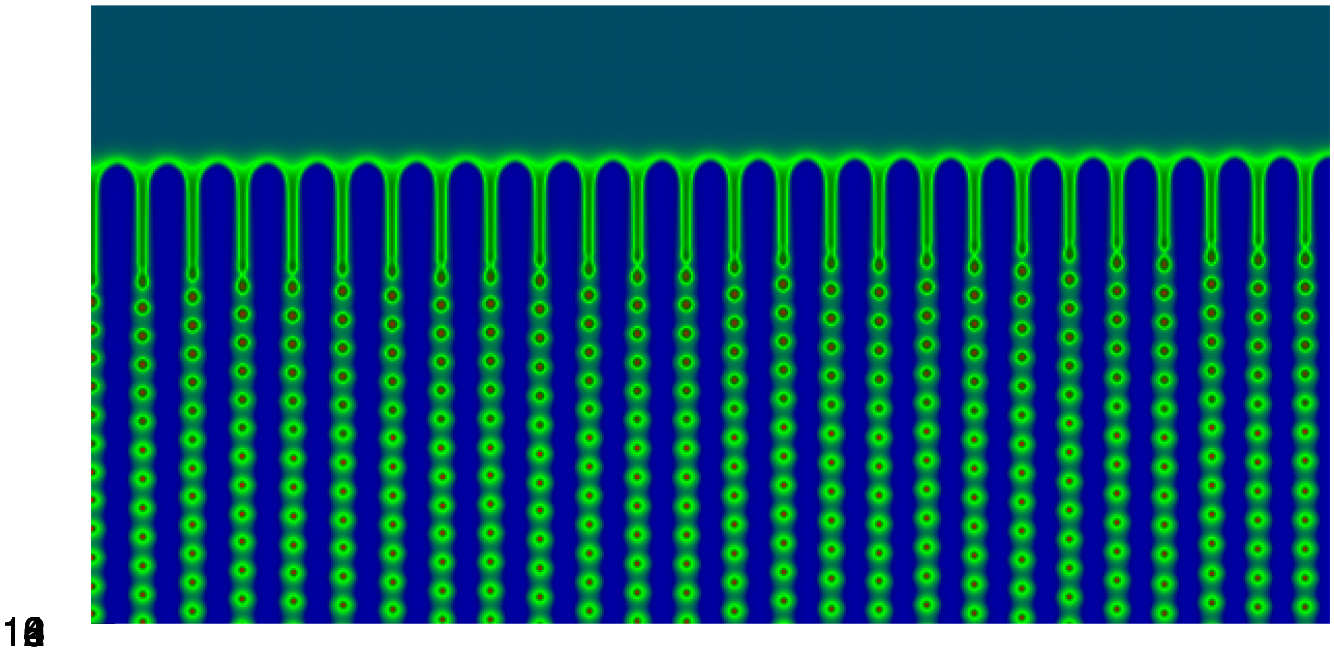}
\caption{}\label{fig_figure2}
\end{subfigure}
\caption{Steady-state cellular growth fronts for two different solidification conditions, are presented using Nb composition map. Only growth fronts are shown here which have been cut off the simulation domain. Color-bar represents the mass fraction of Nb. Growth direction ($y$) is vertical. Three features are seen in these patterns: constant spacing beween the cells, composition variation across the cells, and Nb-enriched droplet formations at the bottom of cell grooves. Details of these features are explained in the text. In (a) $\dot{T}$ = $3 \times 10^5$ K s$^{-1}$, PDAS = \SI{0.71}{\micro\metre} and in (b) $\dot{T}$ = $10^6$ K s$^{-1}$, PDAS = \SI{0.23}{\micro\metre}. Cells are coarser in (a) than (b) due to higher cooling rate. Note that Nb concentration variations along lines A, B and C from (a) are used to study microsegregation in sec.~\ref{sec_solute}.}\label{fig_figure}
\end{figure}

The essence of cellular solidification from the phase-field simulations is as follows. After an initial transient, Mullins-Sekerka instability~\cite{Mullins1964} develops rapidly in the solid-liquid interfaces resulting in the onset of cellular structures. There exist dynamic events like cell merging/splitting at the intermediate stages of growth; however, after a certain stage, the number of cells appearing in the simulation box remains the same and their large-scale geometries (tip and trunk) do not evolve any more. At this stage, the cell tips grow at a constant velocity equal to the solidification velocity and the temperature also remains constant at the tips. This is called steady state. In figure~\ref{fig_figure}, we present the simulated cellular patterns at this stage. Three essential features can be seen in these structures. First, the average distance between the tips of neighboring $\gamma$-cells remains constant, which is documented as the primary dendrite arm spacing (PDAS). This is described in section~\ref{sec_pdas}. Second, there is a significant Nb composition variation in the liquid ahead of the tips, between the cells, and across the solid cells. Niobium is rejected into the liquid by the growing cells and thus the intercellular regions become enriched. This is further described in section~\ref{sec_solute}. Third, circular droplets form in the cell grooves. In steady state, the average distance between the cell tips and the grooves remains constant for a given alloy and processing conditions. To maintain this solidification distance, Nb-rich droplets periodically pinch off from the bottom of the cell grooves resulting in an array of circular pockets. These circular droplets, over time, become highly enriched with solute and, given time, could transform to a secondary phase. However, the dynamics and mechanisms of this are not considered, nor captured in the present theory, since the binary model does not represent any phases beyond L and $\gamma$. Similar features have been reported in experiments as well as in simulations~\cite{ungar1985, Boettinger1988, Boettinger1999,Provatas2017}.

\subsubsection{Dendrite arm spacing}\label{sec_pdas}

The average PDAS from the simulated cellular patterns is estimated by counting the number of cells appearing along the width of the systems (perpendicular to growth direction or $x$ direction) and then dividing by the system width. For the solidification conditions simulated, PDAS obtained are in the range \SI{0.14}{\micro\metre} to \SI{1.6}{\micro\metre}. Interestingly, Amato \emph{et al.}~\cite{Amato2012} have also experimentally observed the cellular/dendritic microstructures with PDAS between \SI{0.5}{\micro\metre} to \SI{1.0}{\micro\metre} in an IN718 alloy, when the same laser processing parameters, as given in table~\ref{table_param_laser}, were used.

\begin{figure}[h]
\centering
\begin{subfigure}[t]{0.6\textwidth}
\centering
\includegraphics[scale=0.26]{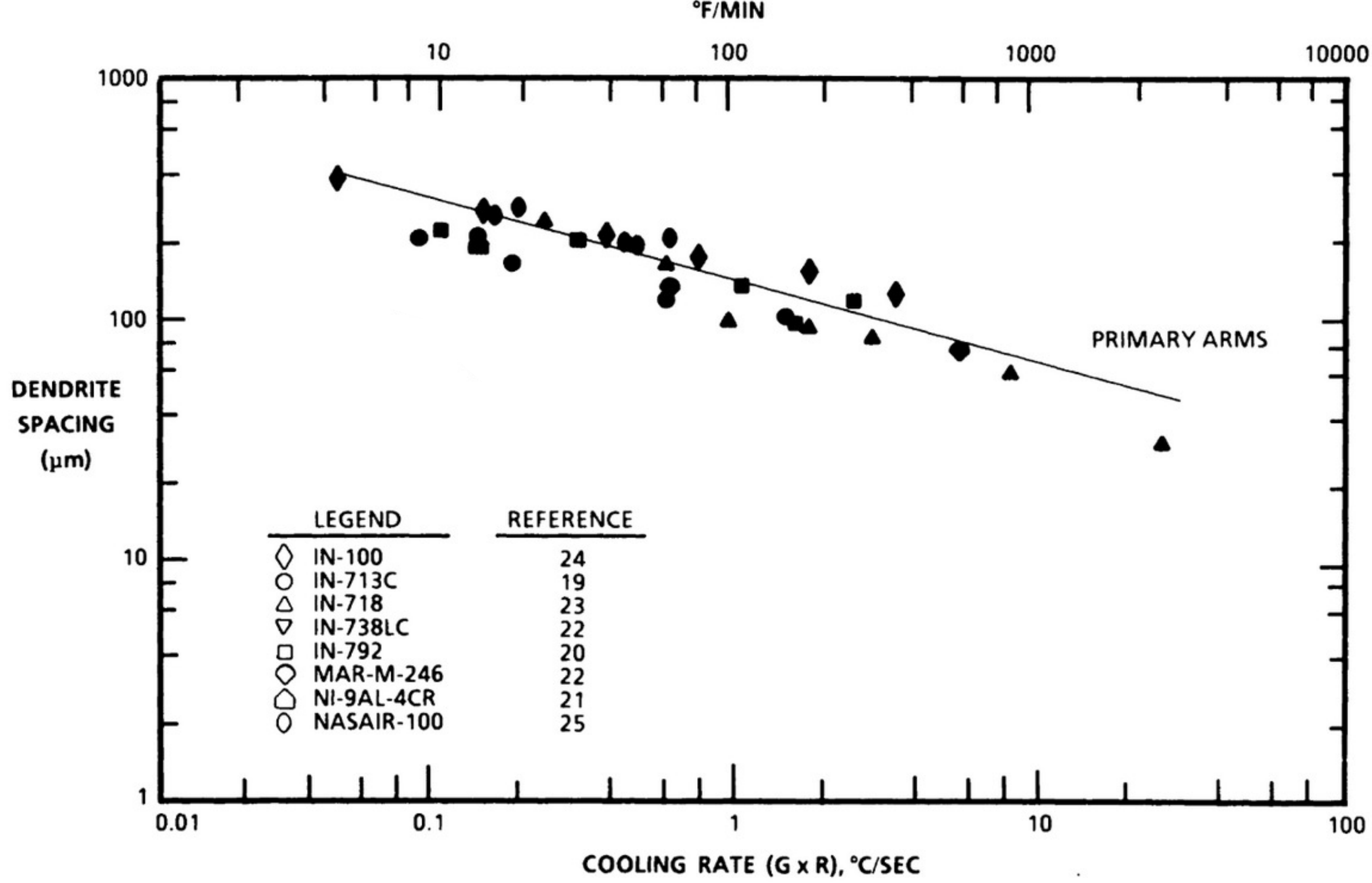} 
\caption{}\label{fig_cooling_data}
\end{subfigure}%
\begin{subfigure}[t]{0.4\textwidth}
\centering
\includegraphics[scale=0.26]{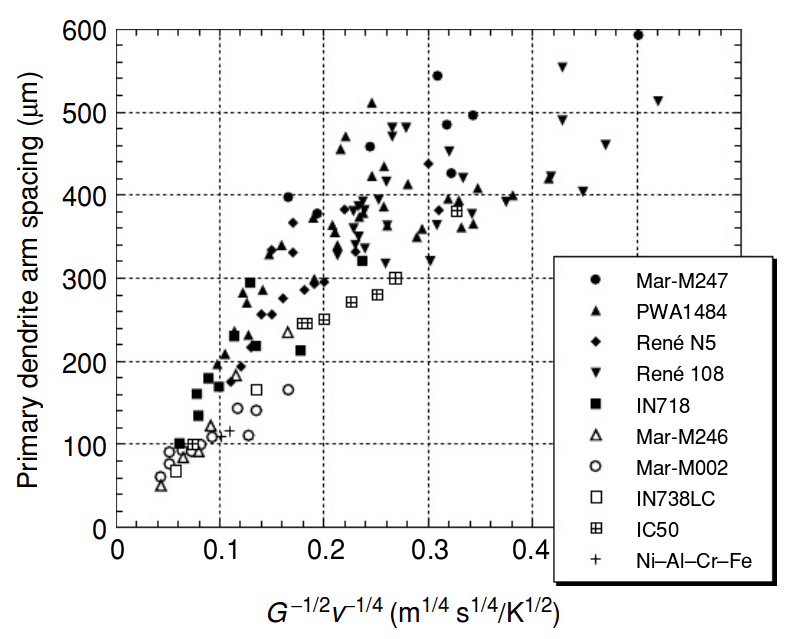}
\caption{}\label{fig_cooling_data2}
\end{subfigure}
\caption{Collection of PDAS data from experiments are presented against cooling rates for different commercial alloys. In (a) $PDAS$ is plotted and fit against $(GV)^n$ in a $\log$ scale (reproduced from~\cite{Bouse1989}, with permission from Elsevier). In (b), $PDAS$ is plotted against $G^{-0.5} V^{-0.25}$ (reproduced from~\cite{Whitesell2000}, with permission from Springer).}\label{fig_data}
\end{figure}

\begin{figure}[h]
\centering
\includegraphics[scale=0.7]{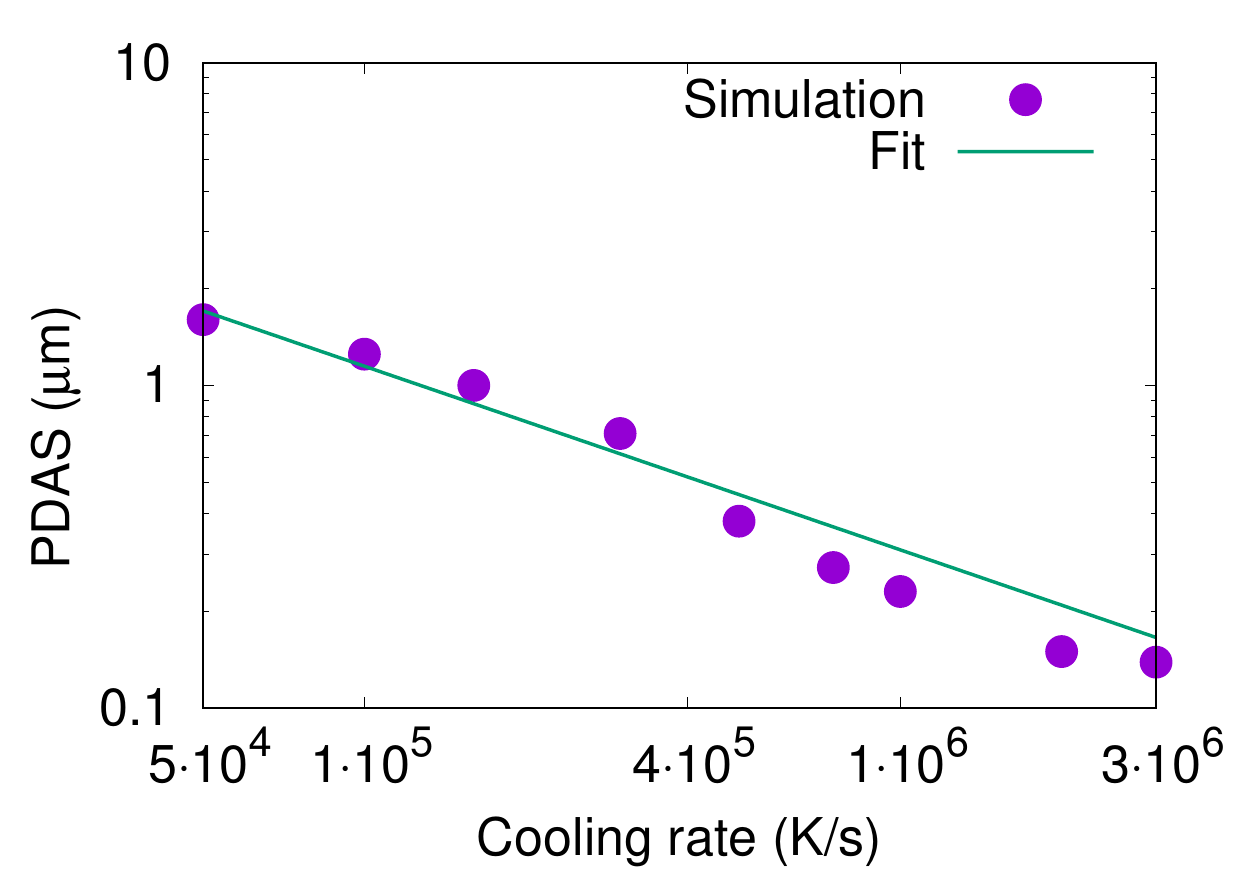}
\caption{Simulated PDAS values are plotted against cooling rates with a fit to equation~\ref{eq_pdas_cooling} in $\log$ scale to obtain $A$ = \SI{800}{\micro\metre}, $n$ = -0.57.}\label{fig_pdas_cooling}
\end{figure}

The simulated results are compared with several theoretical estimates. As mentioned earlier, cell/dendrite spacing in binary alloys under steady-state growth are often characterized by $A G^{m} V^{n}$, where $A$, $m$ and $n$ are constants (refer figures~\ref{fig_cooling_data} and~\ref{fig_cooling_data2}). The simplest form representing this correlation would be a power law with $m = n$, and is given by
\begin{equation}\label{eq_pdas_cooling}
PDAS = A(GV)^n =  A(\dot{T})^n ,
\end{equation}
where $A$ and $n$ are fitting parameters. In figure~\ref{fig_cooling_data}, PDAS vs. $\dot{T}$ data are collected from experiments for different Ni-based superalloys. The approximate $A$ values obtained from these data range between \SI{140}{\micro\metre} to \SI{150}{\micro\metre} and $n$ between -0.4 to -0.5. Note that the maximum cooling rate used in figure~\ref{fig_cooling_data} is 100 K s$^{-1}$. However, the operating cooling rates in the present phase-field simulations are between $5\times10^4$ K s$^{-1}$ to $3\times10^6$ K s$^{-1}$. We plot the simulated PDAS values against the cooling rates in figure~\ref{fig_pdas_cooling}. It is evident that the dendrite arm spacing decreases as the cooling rate increases. The fitting values obtained from this plot are $A$ = \SI{800}{\micro\metre} and $n$ = -0.57.

It should be noted that equation~\ref{eq_pdas_cooling} describes the variation of PDAS with the processing conditions, $G$ and $V$, only. However, in addition to $G$ and $V$, a quantitative description of cells/dendrites needs the integration of the thermophysical parameters as well as the geometry of the evolving structures. Although it is extremely difficult to estimate PDAS for any solidification conditions, there exist a few geometrical models along this line. The present work considers the models of Hunt~\cite{hunt1979} and Kurz and Fisher~\cite{kurz1980}, who applied mass balance and a minimum undercooling criterion at the steady-state cell/dendrite tip to estimate the PDAS. Hunt~\cite{hunt1979} considered only the geometry of the cell/dendrite tip to obtain
\begin{equation}\label{eq_pdas_hunt}
	PDAS =  2.83(k_e \Gamma \Delta T_0 D_l)^{0.25} G^{-0.5}V^{-0.25},
\end{equation}
while Kurz and Fisher~\cite{kurz1980} took the overall geometry (tip and trunk) into account to obtain
\begin{equation}\label{eq_pdas_kf}
	PDAS =  4.3(\Gamma \Delta T_0 D_l/k_e)^{0.25} G^{-0.5}V^{-0.25}.
\end{equation}
Note that these models rely on the proportionality constants $2.83(k_e \Gamma \Delta T_0 D_l)^{0.25}$ or $4.3(\Gamma \Delta T_0 D_l/k_e)^{0.25}$ depending on the geometry of the cellular/dendritic arrays assumed. Both models approximate the cell/dendrite tip using a hemispherical cap while Kurz and Fisher include the additional consideration of the geometry of the trunk, rendering the overall cell/dendrite shape to be an ellipsoid. 

\begin{figure}[h]
\centering
\includegraphics[scale=0.5]{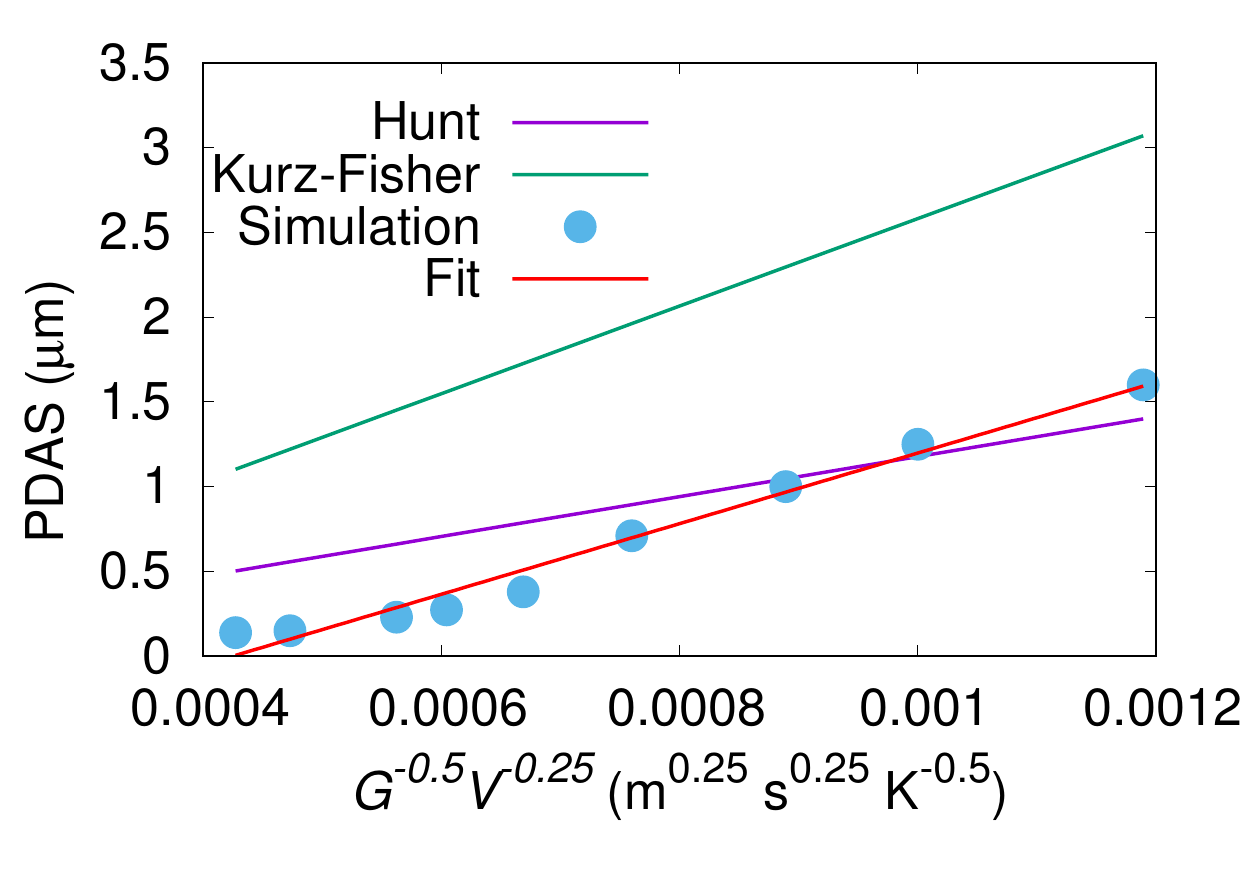}
\caption{Comparison of simulated PDAS results with the geometrical models of Hunt~\cite{hunt1979} (equation~\ref{eq_pdas_hunt}) and Kurz and Fisher~\cite{kurz1980} (equation~\ref{eq_pdas_kf}). While simulated PDAS values are close to the estimates using equation~\ref{eq_pdas_hunt} (Hunt), the slope of the line of best fit using the simulated values is close to $4.3(\Gamma \Delta T_0 D_l/k_e)^{0.25}$ (Kurz and Fisher). The slope of the solid lines are given by $9.56\times10^{-4}$ (Hunt) and $2.6\times10^{-3}$ (Kurz and Fisher). The line of best fit through the simulated values has a slope of $2.1\times10^{-3}$. The slopes are in units of m$^{0.75}$s$^{-0.25}$K$^{0.5}$.}\label{fig_pdas_hunt}
\end{figure}

There is experimental evidence to support the PDAS being proportional to $G^{-0.5}V^{-0.25}$~\cite{Whitesell2000, Ma1998, Quested1984, Bouchard1997} (refer to figure~\ref{fig_cooling_data2}). Following equations~\ref{eq_pdas_hunt} and ~\ref{eq_pdas_kf}, if the theoretical PDAS estimates are plotted against $G^{-0.5}V^{-0.25}$, we obtain the straight lines in figure~\ref{fig_pdas_hunt}, the slopes of which are given by the variable coefficient $2.83(k_e \Gamma \Delta T_0 D_l)^{0.25}$ and $4.3(\Gamma \Delta T_0 D_l/k_e)^{0.25}$. The physical meaning is that these lines represent different cell/dendrite geometries. In this context, the simulated PDAS values are compared with the analytical models in figure~\ref{fig_pdas_hunt}. It is evident that the simulated PDAS do not produce an exact match to the above theories, yet the trend of the results seems to be appropriate. The results are weakly linear with $G^{-0.5}V^{-0.25}$. Interestingly, the slope of the line of best fit through the simulated PDAS values is found to be close to the Kurz and Fisher model. However, there is a nearly constant offset between the values given by the fit and the Kurz and Fisher model. This discrepancy and other variations between the simulation and theory can be attributed to several factors. First of all, the theories are based on a 3-D cell/dendrite geometry, while the simulation results are in 2-D. Moreover, the above theories strictly rely on a particular geometry irrespective of the solidification conditions, but, in reality, solidification geometries change with small modifications in the solidification conditions~\cite{liu1995,kirkaldy1995,ungar1985}. Therefore, the same approximation cannot be applied over the entire regime of $G$ and $V$. 
\subsubsection{Microsegregation}\label{sec_solute}
Niobium concentration variations are investigated in the intercellular regions, the center of the cell tips and deep into the cells. These are represented by lines A, B, and C of figure~\ref{fig_figure1}.  Here, we present results for $G = 10^7$ K m$^{-1}$ and $V$ = 0.03 m s$^{-1}$. It should be noted that the results correspond only to a particular position along the melt pool boundary and thus at a particular depth in the solidified puddle. The trends of the composition variations are found to be similar at other positions.

\begin{figure}[h]
\centering
\begin{subfigure}[t]{0.5\textwidth}
\centering
\includegraphics[scale=0.5]{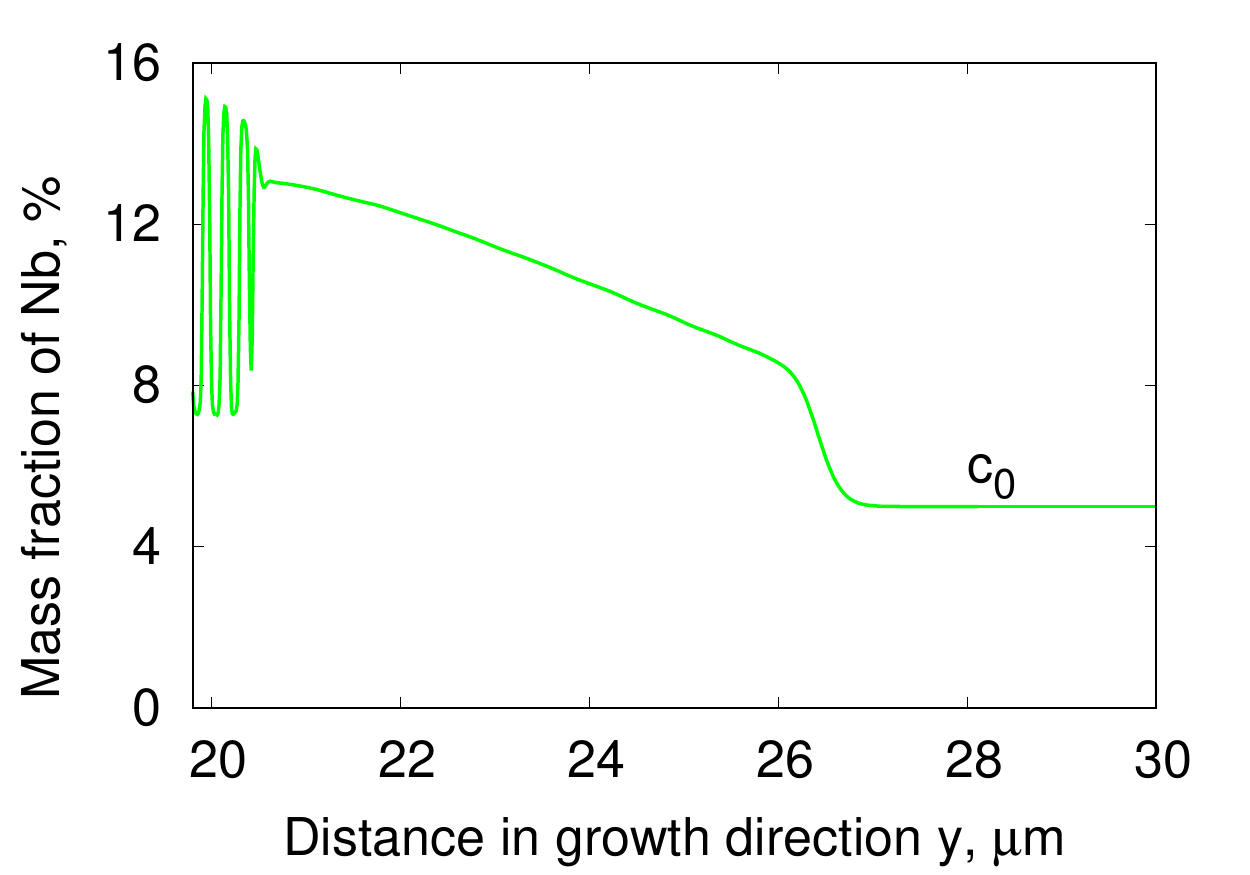}
\caption{}\label{fig_lineA}
\end{subfigure}%
\begin{subfigure}[t]{0.5\textwidth}
\centering
\includegraphics[scale=0.5]{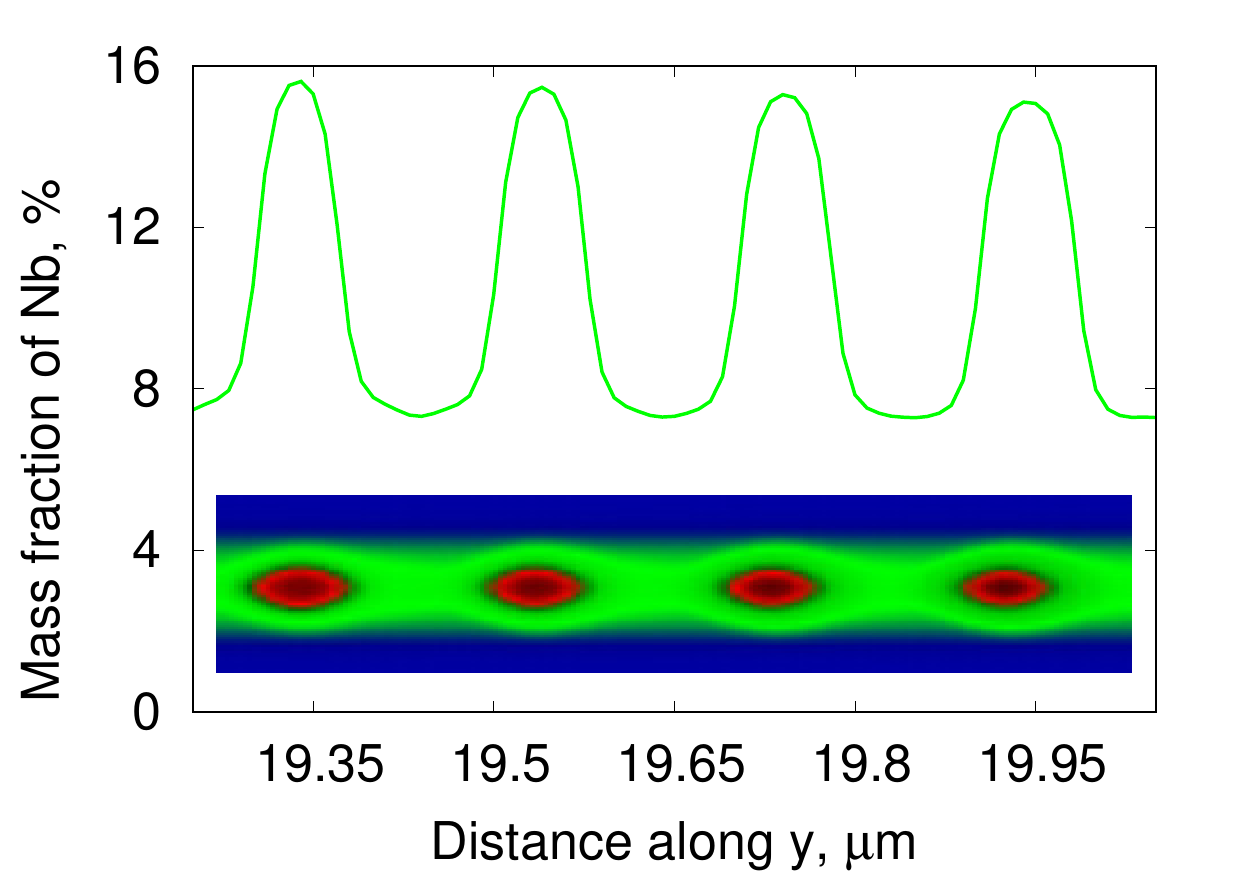}
\caption{}\label{fig_oscillation}
\end{subfigure}
\caption{(a) Nb concentration variations through the intercellular liquid, line A from figure~\ref{fig_figure1}. $c_0$ = \SI{5}{\%} Nb is the far-field liquid concentration. Note that highest Nb compostion inside the liquid droplets, represented by the spikes in the left, is \SI{15.8}{\%}. The Oscillation in concentration in the solid occurs due to pinching off of the liquid from the root of cell grooves at regular time intervals. (b) The oscillation from figure~\ref{fig_lineA} is enlarged in figure~\ref{fig_oscillation} and the representative droplets, from figure~\ref{fig_figure1} line A, are shown in the inset.}\label{fig_droplet_oscillation}
\end{figure}

\begin{figure}[h]
\centering
\includegraphics[scale=0.5]{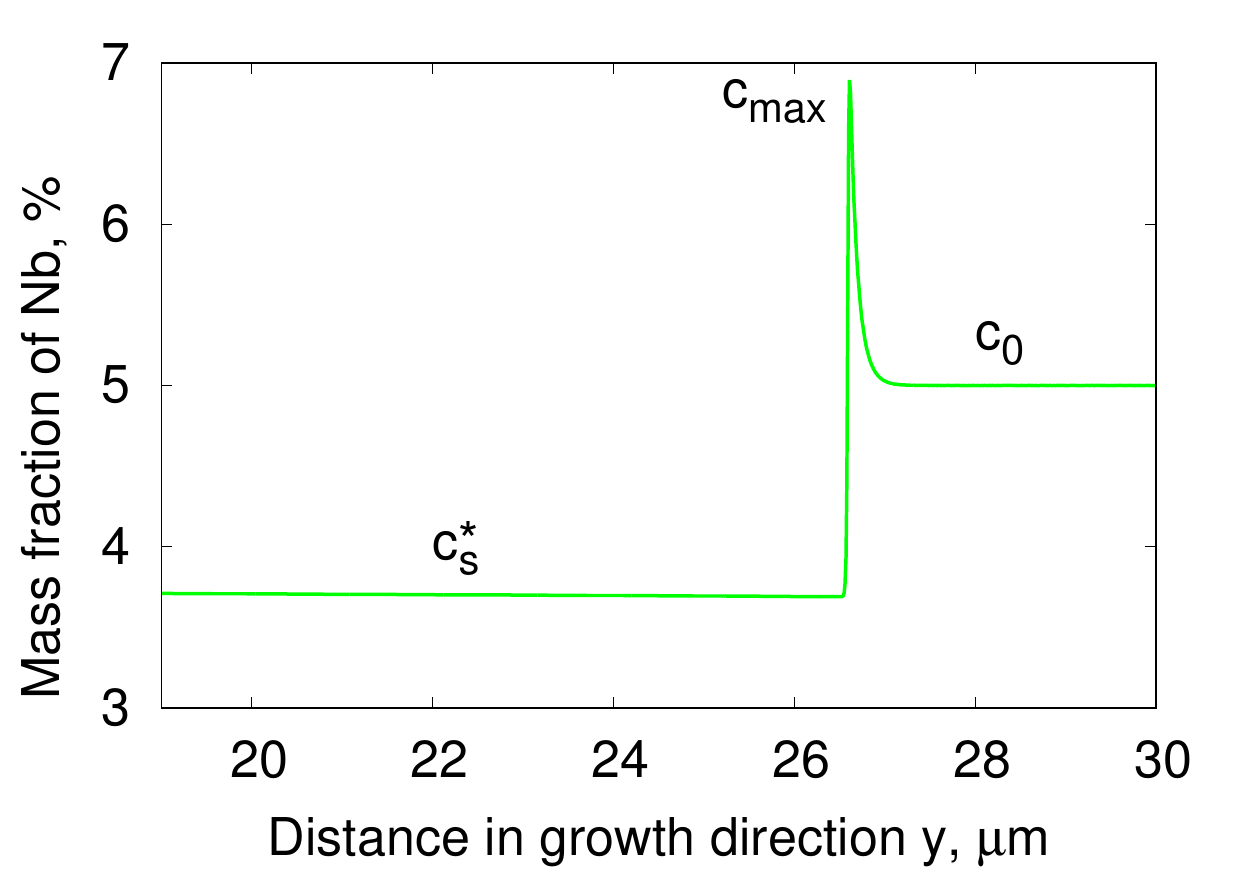}
\caption{Nb concentration variations through a cell tip, line B from figure~\ref{fig_figure1}. $c_{s}^{*}$ = \SI{3.65}{\%} is the Nb concentration inside the cell. $c_{\mbox{\scriptsize max}}$ = \SI{6.87}{\%} is the maximum Nb content at the interface. $c_0$ = \SI{5}{\%} Nb is the far-field liquid concentration.}\label{fig_lineB}
\end{figure}

\begin{figure}[h]
\centering
\includegraphics[scale=0.5]{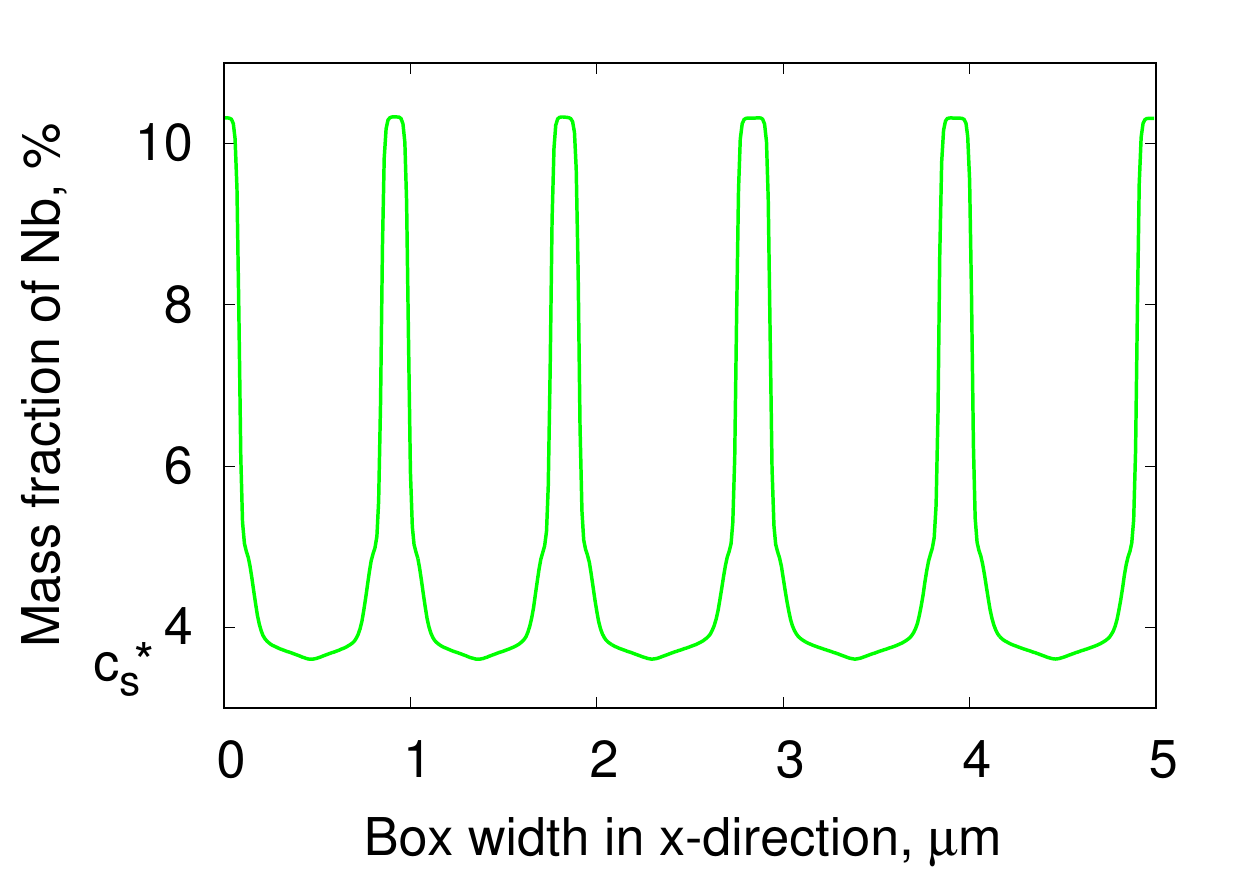}
\caption{Nb concentration variations across the cells, line C from figure~\ref{fig_figure1}. $c_{s}^{*}$ is the Nb concentration inside the cells. For details, please refer to the text.}\label{fig_lineC}
\end{figure}

In figure~\ref{fig_lineA}, Nb concentration profiles are taken along the intercellular liquid between the primary cells (line A in figure~\ref{fig_figure1}). Niobium concentration decreases linearly with distance in the growth direction $y$, while a steeper decrease is observed near the cell tips following the liquidus temperature. Beyond this, the far-field liquid concentration $c_0$ = \SI{5}{\%} Nb is attained. The slope of the linear part in the composition profile between the cells is calculated to be \SI{0.93}{\%} \SI{}{\micro\metre}$^{-1}$. This concentration gradient in liquid between directionally solidified cells is compared to an analytical solution. If curvature effects are ignored, the liquid composition $c_l$ at any position $y$ behind the cell tips can be obtained by letting $G = \frac{dT}{dy}$ and $m_l = \frac{d T}{d c_l}$, where $T$ is the corresponding isotherm temperature at $y$. To calculate the composition gradient in the $y$ direction, $G$ and $m_l$ are combined to obtain $\frac{d c_l}{dy} = \frac{G}{m_l}$. This yields \SI{0.95}{\%} \SI{}{\micro\metre}$^{-1}$, which agrees well with the simulated value.

In figure~\ref{fig_lineB}, Nb concentration is presented through the center of the cells into the liquid in the growth direction (line B in figure~\ref{fig_figure1}). Composition remains almost constant from the cell core to the tip region as the diffusivity of solute in the solid is extremely small. We will refer to this as the cell tip solid concentration $c_{s}^{*}$ in the remaining text. A spike is seen in the Nb composition at the liquid side of the interface due to the rejection of Nb by the growing cells. Beyond this, composition decays rapidly and eventually reaches $c_0$ far into the liquid. Note that for a planar interface in local equilibrium, the maximum Nb content in the liquid and solid side of the interface is given by $c_0/k_e$ = \SI{10}{\%} Nb and $c_0$ = \SI{5}{\%} Nb, respectively (refer to the red curve in figure~\ref{fig_diffusion_1d}). However, for a cell, the details of the diffusion at the tips is different compared to the plane front solidification. This results in different compositions at the cell interface, $c_{\mbox{\scriptsize max}}$ = \SI{6.87}{\%} Nb in the liquid and  $c_{s}^{*}$ = \SI{3.65}{\%} Nb in the solid. This lower amount of Nb in the solid represents solute depletion in the core of the cell. We further note that the ratio of $c_{s}^{*}/c_{\mbox{\scriptsize max}}$ =  0.53, whereas $k_e = 0.48$. Interface partitioning effect is further analyzed in section~\ref{sec_microsegregation}.

\begin{figure}[h]
\centering
\includegraphics[scale=0.5]{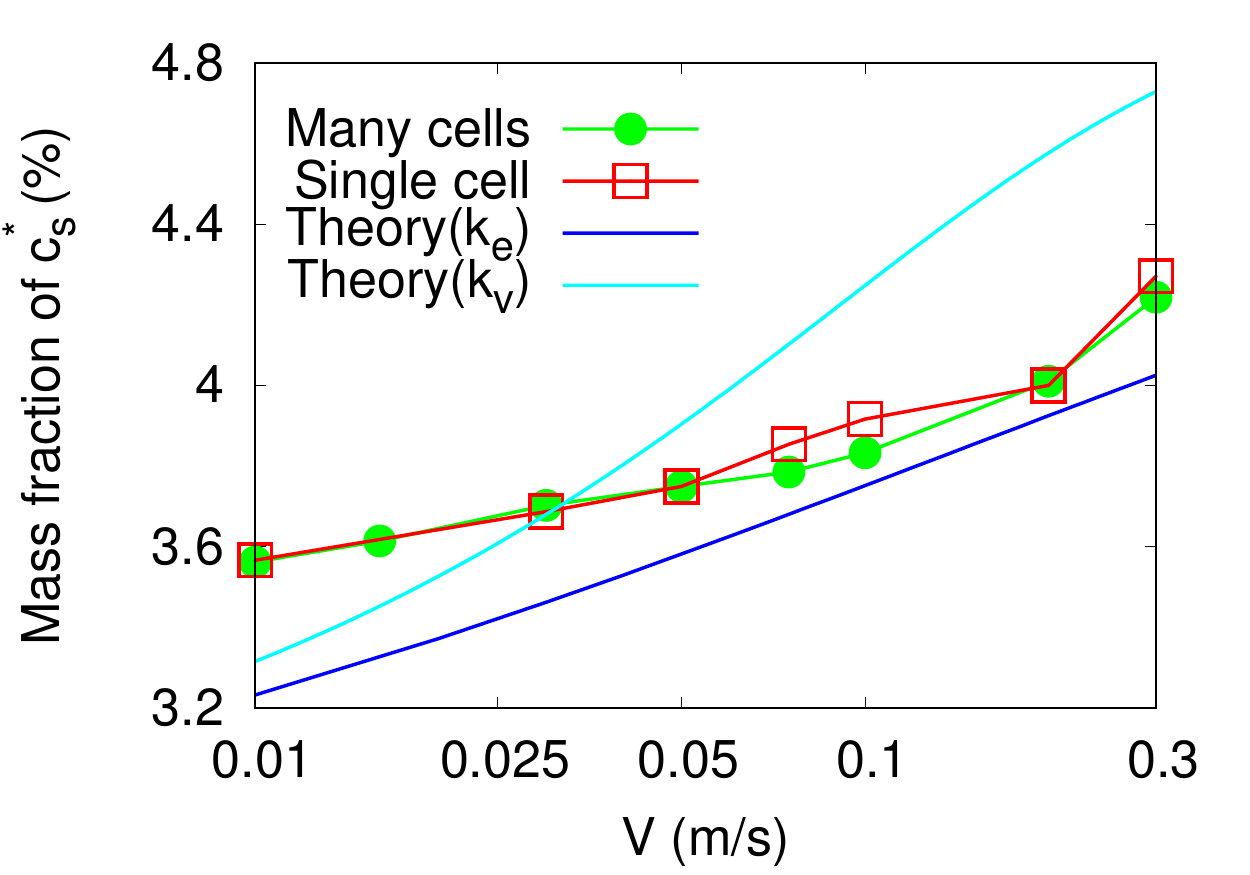}
\caption{Simulated $c_{s}^{*}$ values are compared with the theory (equation~\ref{eq_iv}) for equilibrium $k_e$. Data from many cell simulations are represented with circles and single cell with squares. The comparison with $k_v$ is needed for future reference and is explained later in the text.}\label{fig_iv}
\end{figure}

In figure~\ref{fig_lineC}, Nb profiles are taken along an isotherm that is normal to the growth direction and deep into the mushy zone (line C in figure~\ref{fig_figure1}). The average composition across line C is equal to $c_0$ as it should be for steady state solidification. Here, the compositions at the bottom and the top of the U-shaped profiles correspond to solid and liquid compositions at the symmetrical centers of the cell core and the intercellular region, respectively. Note that $c_{s}^{*}$ in figure~\ref{fig_lineB} is reflected in the lowest values of figure~\ref{fig_lineC}.
Following Kurz and Fisher~\cite{kurzbook}, $c_{s}^{*}$ can be estimated by a mathematical analysis of the diffusion fields around an isolated paraboloid cell/dendrite tip leading to the following growth equations:
\numparts
\begin{eqnarray}
G_c &=& -\frac{V}{D_l}k_ec_{s}^{*}(1-k_e), \label{eq_iv_first} \\
R &=& 2\pi\left[ \frac{\Gamma}{m_l G_c - G} \right]^{\frac{1}{2}}, \\
c_{s}^{*} &=& \frac{k_e c_0}{1-(1-k_e)Iv(P)}, \label{eq_iv} \\
Iv(P) &\equiv& P \exp(P)E_1(P), \\
P &=& \frac{RV}{2D_l}. \label{eq_iv_last} 
\end{eqnarray}
\endnumparts
$G_c$ is the composition gradient in the liquid, $R$ is the cell/dendrite tip radius and $E_1(P)$ is the first exponential integral of the cell/dendrite P\'{e}clet number $P$. Referring to equations~\ref{eq_iv_first}-\ref{eq_iv_last}, $c_{s}^{*}$ values are solved for the same thermophysical and processing parameters used in the phase-field simulations. The estimated $c_{s}^{*}$ values are then compared with the simulated $c_{s}^{*}$ values in figure~\ref{fig_iv}. The simulated $c_{s}^{*}$ values are reasonably close to the estimations. However, certain differences are seen between the simulated and estimated values. The theory is based on an isolated cell/dendrite tip and hence does not consider the interactions between neighboring cells/dendrites, nor is the surface tension anisotropy considered. In order to illustrate the interactions between multiple cells, results are compared with simulations for a single steady state cell in figure~\ref{fig_iv}. However, the differences between the data for many cells vs. single cell are negligible in the present scenario. Therefore the differences between data and theory can be expected due to a reduced geometry (2-D rather than 3-D) and the capillary effects which are significantly different in 3-D~\cite{Rappazbook,Wang2003,Lee2010}. The anti-trapping current also has an effect on the above discrepancies which is explained in section~\ref{sec_microsegregation}. Note that $c_{s}^{*}$ increases with increasing $V$. This behavior is expected when the growth velocity approaches $V_{ab}$.

\begin{figure}[h]
\centering
\includegraphics[scale=0.5]{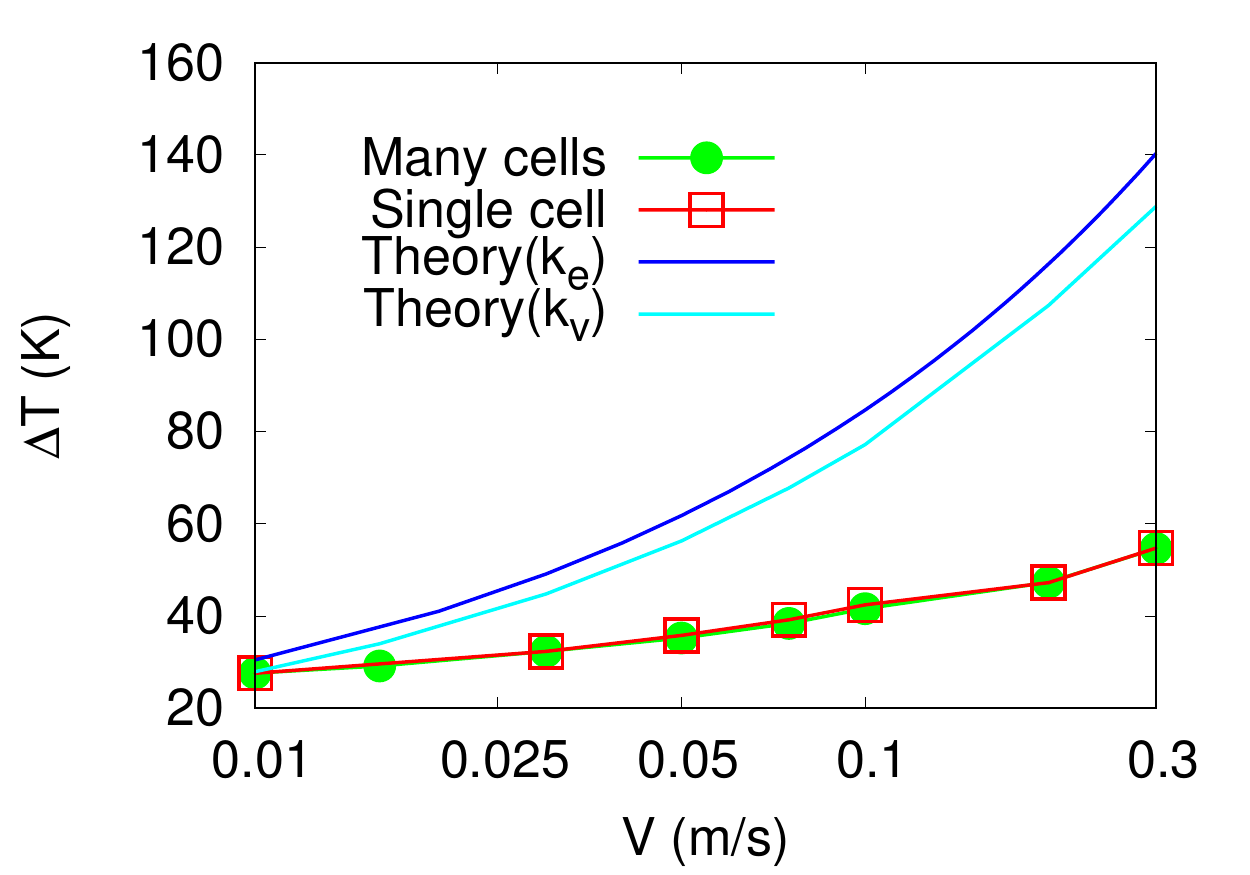}
\caption{Cell tip undercooling $\Delta T$ increases with the solidification speed $V$. Simulation results are compared with theoretical estimates for constant $k_e$. Data from many cell simulations are represented with circles and single cell with squares. The comparison with $k_v$ is needed for future reference and is explained later in the text.}\label{fig_undercooling}
\end{figure}

Equations~\ref{eq_iv_first}-\ref{eq_iv_last} can also be used to determine the cell/dendrite tip temperature or the tip undercooling $\Delta T$ below the liquidus temperature in the following way~\cite{kurzbook}:
\begin{eqnarray}~\label{eq_undercooling}
\Delta T &=& \Delta T_{solutal} + \Delta T_{curvature} \nonumber \\
&=& 2m_l c_0 P (1-k_e) + 2\Gamma/R . 
\end{eqnarray}    
The total undercooling at the tip is due to its composition and curvature, given by $\Delta T_{solutal}$ and $\Delta T_{curvature}$ in equation~\ref{eq_undercooling}, with the former being dominant. For comparison, the average $\Delta T$ values are obtained from the simulated steady-state cell tips and are plotted against $V$ in figure~\ref{fig_undercooling}. Following equations~\ref{eq_iv_last} and~\ref{eq_undercooling}, $\Delta T$ increases with increasing $V$ and certain differences are seen between the theoretical and simulated values. The reasons for the differences are the same as described for figure~\ref{fig_iv}. Similar behavior has been reported in~\cite{Boettinger1999,amberg2008,Lee2010,Kundin2015}.

\subsubsection{Interface partitioning}\label{sec_microsegregation}
\begin{figure}[h]
\centering
\includegraphics[scale=0.6]{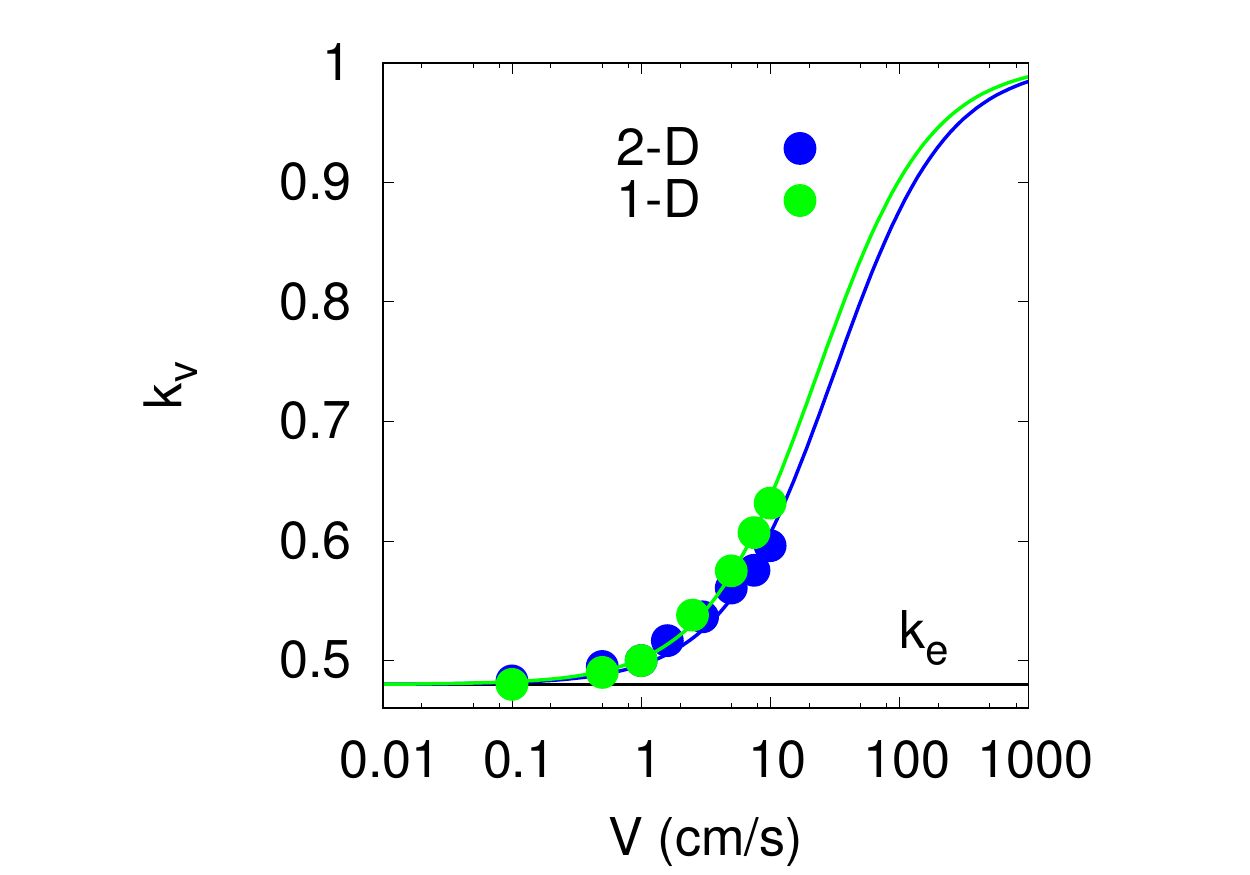}
\caption{Simulated $k_v$ from 2-D and 1-D simulations are plotted for various growth rates along with a fit to equation~\ref{eq_aziz}. Note that with increasing $V$, $k_v$ increasingly deviates from equilibrium $k_e$.}\label{fig_vd_2d}
\end{figure}

As discussed in the above section, Nb composition in the solid and liquid at the interface do not obey local equilibrium. To quantify this departure from local equilibrium, the degree of Nb partitioning at the tips is considered. The ratio of Nb composition in the cell (just behind the tips) to the maximum Nb composition in the liquid is taken as a measure of the velocity-dependent partition coefficient $k_v$, which is expressed as~\cite{Boettinger1999, Danilov2006, Wheeler1993}
\begin{equation}\label{eq_kv}
k_v = \frac{c_{s}^{*}}{c_{\mbox{\scriptsize max}}}.
\end{equation}
We calculate the $k_v$ values from the Nb compositions of the steady-state cell tips for different $V$. This is shown in figure~\ref{fig_vd_2d}. Clearly, $k_v$ increases with increasing $V$. This occurs even though the phase-field simulations are conducted with the equilibrium value of $k_e$ and the anti-trapping current~\cite{Echebarria2004} has been added to correct deviations from local interface equilibrium. However, the anti-trapping current used is only accurate to second order in the small parameter $W_0/d_0$ in the thin-interface asymptotic analysis. The model requires improvement if local interface equilibrium is required at these high solidification velocities. On the other hand, some level of solute trapping should be present. The physical meaning is that with increasing interface velocity $V$, solute has less time to redistribute within the interface. The simulated trapping behavior can be characterized by the Aziz trapping function, which relates $V$ with $k_v$~\cite{Aziz1982}
\begin{equation}\label{eq_aziz}
k_v (V) = \frac{k_e + V/V_D}{1 + V/V_D}.
\end{equation}
$V_D$ is the interface diffusion speed, which is given by $D_l/a_0$ with $a_0$ the characteristic interface width on the order of interatomic distances. In figure~\ref{fig_vd_2d}, the simulated $k_v$ values are fit with the Aziz trapping function (equation~\ref{eq_aziz}) for a value of $V_D$ = 0.31 m s$^{-1}$. It should be noted that $V_D$ is a characteristic velocity for a given alloy related to the magnitude of its solute-trapping behavior. To our knowledge, no experimental data is available for correct value of $V_D$ for Ni-Nb.

It is interesting to note that $k_v = 1$ in figure~\ref{fig_vd_2d} signifies zero partitioning of the solute, or complete solute-trapping, which can be attained for $V$ greater than 10 m s$^{-1}$. However, this is far beyond the $V$ used in the present work. For our intermediate $V$, partial partitioning results in $k_e < k_v <1$. For a velocity of 0.001 m s$^{-1}$  or below, equilibrium partitioning is recovered with the constant $k_v = k_e$. 

\begin{figure}[h]
\centering
\includegraphics[scale=0.5]{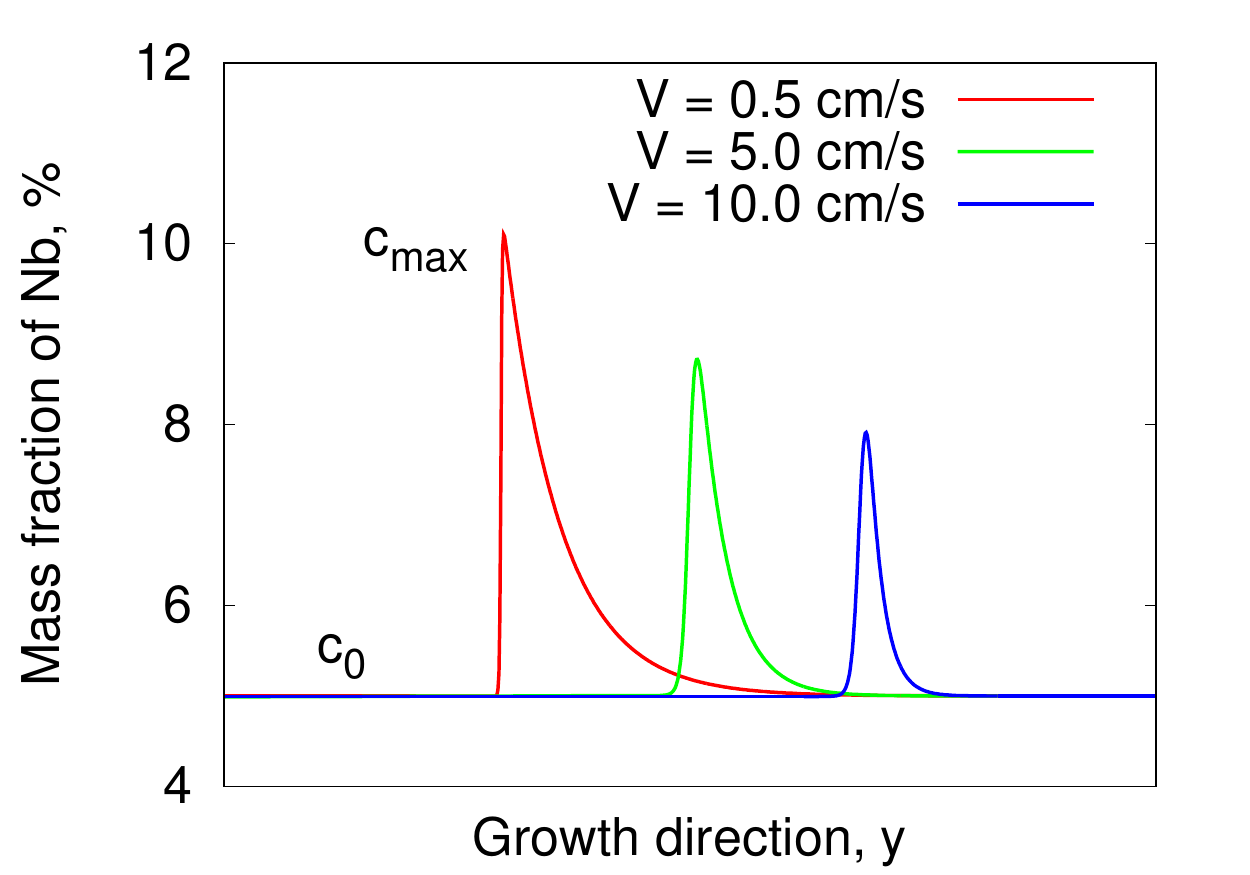}
\caption{Nb concentration profiles developed for different $V$ for planar front growth in 1-D. With increasing $V$, solute partitioning at the interface decreases leading to decreasing $c_{\mbox{\scriptsize max}}$.}\label{fig_diffusion_1d}
\end{figure}

In order to document the solute-trapping in the present model in the most simple way, 1-D simulations are performed under the same conditions as in 2-D, to render the Nb partitioning in a planar front growth mode. This is different from the 2-D simulations in that far-field liquid and solid compositions are the same (refer to figure~\ref{fig_diffusion_1d}). The maximum Nb composition $c_{\mbox{\scriptsize max}}$ at the interface decreases with the growth speed $V$. Fitting of equation~\ref{eq_aziz} to the 1-D simulation results, with $k_v$ $ = c_0/c_{\mbox{\scriptsize max}}$, yields $V_D$ = 0.23 m s$^{-1}$. The difference in $V_D$ between 1-D and 2-D may be due to curvature effects at the cell tips, which in turn depend on the solidification velocity $V$. Thus $R$ becomes a function of $V$, i.e., $R(V)$. The curvature-corrected partition coefficient can be represented as~\cite{Karma2001, Mullis2010}
\begin{equation}
k_v (R(V)) = \frac{c_0}{c_{\mbox{\scriptsize max}}(V)} \left(1-(1-k_e)\frac{d_0}{R(V)}\right).
\end{equation}   

One phenomenological approach to obtaining a predetermined level of solute-trapping in the phase-field model is as follows. The equilibrium solidification parameters are replaced with the velocity-dependent parameters to reflect the non-equilibrium changes at the interface. The effect of the velocity-dependent partition coefficient $k_v$ is given by equation~\ref{eq_aziz}. Similarly, the liquidus slope $m_l$ becomes a function of $V$ and takes the form~\cite{kurzbook}
\begin{equation}
m_v (V) = m_l \frac{1-k_v\left[1-\ln(k_v/k_e)\right]}{1-k_e}.
\end{equation} 
For evaluating the solute concentrations at the cell tips, velocity-dependence is also introduced into equation~\ref{eq_iv}, resulting in~\cite{kurzbook}
\begin{equation}
c_{s}^{*}(V) = \frac{k_v c_0}{1-(1-k_v)Iv(P)}.
\end{equation} 

Referring to equations~\ref{eq_iv_first}-\ref{eq_iv_last}, $c_{s}^{*}(V)$ values are solved in the same way but now with the modified $k_v$ and $m_v$. The comparison with the simulation is given in figure~\ref{fig_iv}. At a lower $V$, this theory provides somewhat better estimates. However, for high $V$, the convergence is not satisfactory, as the theory overestimates. Moreover, in terms of undercooling (refer to figure~\ref{fig_undercooling}), the velocity-dependence estimates are closer to the simulation results. Nevertheless, the comparison is not satisfactory. To explain this behavior quantitatively, we draw our attention to the phase-field model components, particularly the anti-trapping current in equation~\ref{eq_c}. The purpose of this is to eliminate artificial solute-trapping due to the thick diffuse interfaces used for the phase-field simulations. Moreover, this additional solute flux is meant to correct for behavior in the low solidification velocity regime. The effects of this current toward the physical solute-trapping in the rapid solidification regime is not clear at the moment and work is under way to understand this well.

Moreover, the present study does not consider the effects of convection on the primary arm spacing and Nb segregation. Effects of convection on the primary arm spacing is not as pronounced as compared to the secondary and tertiary arms, which are not observed in our simulations, where the solute field interacts in a complex manner with the microstructure~\cite{Wang2003, Lee2010}. However, it is seen that this results in faster growth of the cells as the solute transport is enhanced, leading to solute gradient dips at the tips~\cite{Wang2003, Lee2010}. 

The primary arm spacings simulated in the present work are smaller than \SI{1.6}{\micro\metre}. Such dense cellular structures provide significant resistance to the flow following an exponential increase of the damping effect in the mushy region and hence the reduced effects of convection~\cite{Yang2001,Tan2011}. In addition, consideration of a dilute alloy reduces the convection effects on the compositions~\cite{Wang2003}. Therefore, simulations have been performed with a reasonable approximation for the average behavior of a primary cell/dendrite towards the selection of spacing and the microsegregation patterns.
\section{Summary and Outlook}\label{sec_summary}
We use a macroscopic heat transfer model to obtain melt pool solidification conditions relevant to additive manufacturing. A mesoscopic phase-field model is then used to simulate the cellular patterns formed during solidification under these conditions. Cellular solidification is discussed in regard to primary arm spacing and microsegregation. Primary arm spacing decreases from \SI{1.6}{\micro\metre} to \SI{0.14}{\micro\metre} as the cooling rate increases from $5\times10^4$ K s$^{-1}$ to $3\times10^6$ K s$^{-1}$. Simulated spacings are compared against the geometrical models of Hunt and Kurz and Fisher with reasonable agreements. At steady-state, solute concentrations at the cell tips are found to be smaller than planar front and compare reasonably with an isolated cell/dendrite growth model. Growth velocity-dependence is considered for the alloy parameters to measure the extent of solute-trapping behavior during non-equilibrium solidification. The results indicate that the anti-trapping current within a phase-field model is not adequate to eliminate solute-trapping at these high solidification rates.

The simulated microstructures and the concentration fields can be used as inputs for the simulation of subsequent heat treatment; notably secondary phase formations in between the $\gamma$-cells may arise because the Nb-rich droplets are expected to transform into Laves phases during the subsequent stages of solidification in IN718~\cite{nie2014,Kundin2015}. Particularly for the circular droplets that emerge as a function of pinching off from the cellular structures, a quantitative analysis of these structures can offer insight into the size and distribution of secondary phases. We do not include the effects of fluid flow, Marangoni surface flow and other hydrodynamic effects in the melt pool, each of which may bring significant differences~\cite{Lee2014,Khairallah2016,King2015}. Simulations in 3-D can be performed to explore another degree of freedom in the solute segregation and droplet formations. Finally, we note that the solute-trapping behavior exhibited in the present model needs to be explored. The role of the anti-trapping current, which is designed to capture local interfacial equilibrium properties in the low velocity regime for large diffuse interfaces, needs to be better understood in this intermediate to high velocity regime. A quantitative understanding of this mechanism at rapid solidification regime may lead to a more accurate, physical description of solute-trapping.

\section*{Acknowledgements}
We thank William J. Boettinger for careful reading of the manuscript and for many fruitful discussions. S.G. thanks Trevor Keller for his help in constructive feedback and high performance computing. S.G. acknowledges Lyle Levine and Eric Lass for interactions on the experimental results at NIST that inspired the present work. N.O.-O. acknowledges the following financial assistance: Award No. 70NANB14H012 from U.S. Department of Commerce, National Institute of Standards and Technology as part of the Center for Hierarchical Materials Design (CHiMaD).


\section*{References}
\bibliographystyle{unsrt}
\bibliography{papers}
\end{document}